\headline={\ifnum\pageno>1 \hss \number\pageno\ \hss \else\hfill \fi}
\pageno=1
\nopagenumbers
\hbadness=1000000
\vbadness=1000000

\voffset=0.8in
\centerline{\bf SUMMING OVER THE WEYL GROUPS OF $E_7$ AND $E_8$ }
\vskip 15mm
\centerline{\bf H. R. Karadayi \footnote{*}{
e-mail: karadayi@itu.edu.tr} and M.Gungormez}
\centerline{Dept.Physics, Fac. Science, Tech.Univ.Istanbul }
\centerline{ 80626, Maslak, Istanbul, Turkey }
\vskip 20mm
\centerline{\bf{Abstract}}
\vskip 1mm

It is known that summations over Weyl groups of Lie algebras is a
problem which enters in many areas of physics as well as in mathematics.
For this, a method which we would like to call {\bf permutation weights}
has been previously proposed for pairs $(G_N , A_{N-1})$ of Lie algebras.
It is now extended for $(E_7 , A_7)$ and also $(E_8 , A_8)$.
It is clear that these are the most non-trivial ones and hence deserve
studying separately.

In order to obtain the results of these summations in practice, it is shown
that some simplifications occur in the method which is previously proposed
for pairs $(A_N , A_{N-1})$ in an unpublished work.

\vskip 15mm
\vskip 15mm
\vskip 15mm
\vskip 5cm
\hfill\eject

\voffset=0in

\vskip 3mm
\noindent {\bf{I.\ INTRODUCTION}}
\vskip 3mm

In a previous paper {\bf [1]}, we have shown that the method of the so-called
permutation weights makes possible explicit applications of character
formulas in a manner which is independent from the rank or complexity of
the underlying Lie algebra $G_N$ of rank N. The proposed method works
in terms of the sub-algebra $A_{N-1}$ of $G_N$. Simplification comes out
in summations over the Weyl group $W(G_N)$. The main emphasis is on the
following two facts:

${\bf \bullet}$ for any finite Lie algebra $G_N$, the dimensions of Weyl orbits
$W(\Lambda^{++})$ of strictly dominant weights $\Lambda^{++}$ are all the same ${\bf \bullet}$

${\bf \bullet \bullet}$ permutation weights form a sub-set $\varphi(\Lambda^{++})$ of
$W(\Lambda^{++})$ and their dimensions dim$\varphi(\Lambda^{++})$ are
independent from $\Lambda^{++}$, namely
dim$\varphi(\Lambda^{++})$ = dimW($G_N$)/dimW($A_{N-1}$) ${\bf \bullet \bullet}$

\noindent The problem seems to be quite involved and complex when one recalls
dim$\varphi(\Lambda^{++})$ =576 and 17280 for $E_7$ and $E_8$ respectively
though it will be more tractable thanks to the fact that $E_7$ and $E_8$ have
the sub-groups $A_7$ and $A_8$. Note here that
dim$\varphi(\Lambda^{++})$ =72 and 1920 for pairs ($E_7 , A_7$) and
($E_8 , A_8$). To this end, let us remark that permutation weights turn out
to be dominant weights. All these will be investigated in the
following two sections and it will be seen that they allow us to obtain
explicit results for summations over 2903040 and 696729600 elements of the
Weyl groups of $E_7$ and $E_8$.

One must however warn the reader that the problem is still not sufficiently
tractable in the absence of some simplifications in the method for pairs
($A_N , A_{N-1}$). It will therefore be shown in the last section that,
for pairs ($A_N , A_{N-1}$), the method of permutation weights seems to be
more simple in an appropriate specialization of formal exponentials. To this
end, the introduction of a reduction formula governing degenerated
Schur functions must also be noted.

We refer to the excellent book of Humphreys {\bf [2]} for basic knowledge on
Lie algebra technology. We normalize ourselves as being in line with the
following Coxeter-Dynkin diagrams

$$ \hskip1.4cm  8  $$
$$  1 \hskip0.5cm 2 \hskip0.5cm 3 \hskip0.5cm 4 \hskip0.5cm 5 \hskip0.5cm 6 \hskip0.5cm 7 $$

\noindent for $E_8$,

$$  1 \hskip0.5cm 2 \hskip0.5cm 3 \hskip0.5cm 4 \hskip0.5cm 5 \hskip0.5cm 6 \hskip0.5cm 7 \hskip0.5cm 8 $$

\noindent for $A_8$,

$$ \hskip0.7cm  7  $$
$$  1 \hskip0.5cm 2 \hskip0.5cm 3 \hskip0.5cm 4 \hskip0.5cm 5 \hskip0.5cm 6  $$

\noindent for $E_7$,

$$  1 \hskip0.5cm 2 \hskip0.5cm 3 \hskip0.5cm 4 \hskip0.5cm 5 \hskip0.5cm 6 \hskip0.5cm 7 $$
\noindent for $A_7$.

In the common notation for N=7,8, \ fundamental dominant weights are
$\Lambda_i \in E_N$ , $\lambda_i \in A_N$ while simple roots are
$\beta_i \in E_N$ and $\alpha_i \in A_N$. We define {\bf [3]} the
{\bf fundamental weights} $\mu_I$ \ (I=1,2,.. N+1) of $A_N$ by
$$ \eqalign{
\mu_1 &\equiv \lambda_1  \cr
\mu_I &\equiv \mu_{I-1}-\alpha_{I-1} \ \ , \ \  I=2,3,..N+1 }
\eqno(I.1) $$
or, conversely, by
$$ \lambda_i = \sum_{j=1}^i \ \mu_j \ \ , \ \ i=1,2,.. N.
\eqno(I.2) $$
By definition, $A_N$ fundamental weights are constrained by
$$ \sum_{I=1}^{N+1} \ \mu_I \equiv 0
\eqno(I.3) $$
and they provide us a scalar product
$$ (\mu_I,\mu_J) = \delta_{I,J} - {1 \over (N+1)}
\eqno(I.4) $$
by the aid of which we can obtain all the scalar products we need in the
sequel.

\vskip 3mm
\noindent {\bf II.\ ${\bf (E_7 , A_7)}$ }
\vskip 3mm

The starting point here is the following decompositions of the seven
fundamental $E_7$ Weyl orbits $W(\Lambda_i)$ \ (i=1,2,..7) \ in terms of
$A_7$ Weyl orbits:
$$ \eqalign{
W(\Lambda_1) &= W(\lambda_2) \oplus W(\lambda_6)  \cr
W(\Lambda_2) &= W(\lambda_1 + \lambda_3) \oplus
W(\lambda_2 + \lambda_6) \oplus W(\lambda_5 + \lambda_7)  \cr
W(\Lambda_3) &= W(2 \ \lambda_3) \oplus
W(2 \ \lambda_1 + \lambda_4) \oplus  W(2 \ \lambda_5) \oplus \cr
& \ \ \ \ W(\lambda_1 + \lambda_3 + \lambda_6) \oplus
W(\lambda_2 + \lambda_5 + \lambda_7) \oplus W(\lambda_4 + 2 \ \lambda_7)  \cr
W(\Lambda_4) &= W(2 \ \lambda_2 + \lambda_4) \oplus
W(3 \ \lambda_1 + \lambda_5) \oplus W(\lambda_2 + 2 \lambda_5) \oplus
W(2 \ \lambda_3 + \lambda_6) \oplus \cr
& \ \ \ \ W(2 \ \lambda_1 + \lambda_4 + \lambda_6) \oplus
W(\lambda_4 + 2 \ \lambda_6) \oplus
W(\lambda_1 + 2 \ \lambda_2 + \lambda_7) \oplus
W(\lambda_1 + \lambda_3 + \lambda_5 + \lambda_7) \oplus \cr
& \ \ \ \ W(\lambda_1 + 2 \ \lambda_6 + \lambda_7) \oplus
W(\lambda_2 + \lambda_4 + 2 \ \lambda_7) \oplus
W(\lambda_3 + 3 \ \lambda_7)  \cr
W(\Lambda_5) &= W(2 \ \lambda_2) \oplus W(\lambda_3 + \lambda_5) \oplus
W(2 \ \lambda_1 + \lambda_6) \oplus  W(2 \ \lambda_6) \oplus
W(\lambda_1 + \lambda_4 + \lambda_7) \oplus W(\lambda_2 + 2 \ \lambda_7) \cr
W(\Lambda_6) &= W(\lambda_4) \oplus W(\lambda_1 + \lambda_7)  \cr
W(\Lambda_7) &= W(2 \ \lambda_1) \oplus W(\lambda_1 + \lambda_5) \oplus
W(\lambda_3 + \lambda_7) \oplus W(2 \ \lambda_7)     } \eqno(II.1)    $$
\noindent where $\oplus$ means collection. Note here that, due to the fact
that $E_7$ is a real Lie algebra, its Weyl orbits remain unchanged under
$A_7$ diagram automorphism which are expressed by
$\lambda_i \rightarrow \lambda_{8-i}$. (II.1) allows us to choose
$$ \eqalign{
\Lambda_1 &= \lambda_2             \cr
\Lambda_2 &= \lambda_1+\lambda_3   \cr
\Lambda_3 &= 2 \lambda_3           \cr
\Lambda_4 &= 2 \lambda_3+\lambda_6 \cr
\Lambda_5 &= \lambda_3+\lambda_5   \cr
\Lambda_6 &= \lambda_4             \cr
\Lambda_7 &= \lambda_3+\lambda_7      } \eqno(II.2) $$
\noindent for a correspondence between $E_7$ and its sub-algebra $A_7$.
There could be some choices, in view of (II.1), other than (II.2), the
results will however be the same for any one of them.

As is emphasized above, the permutation weights of the pair $(E_7 , A_7)$
will now be obtained to be $A_7$ dominant weights. All these can be seen
by the use of the method proposed in section II of ref.[1]. We, instead, want
here to follow an alternative way by nominating $E_7$ Weyl group elements
which give us these permutation weights directly. It will be seen that this
will simplify the matter and also makes clear how we assign signatures for
these permutation weights. It is known that the whole Weyl group $W(E_7)$
can be constructed by successive applications of simple Weyl reflections $w_i$.
A Weyl reflection $w_{\beta}$ with respect to any root $\beta$ is known to be
an operation defined on the whole weight lattice of $E_7$ Lie algebra.
For simple roots, $w_i$'s are defined to be the Weyl reflections of length 1,
that is
$$ \ell (w_i) \equiv 1  . \eqno(II.3) $$
A notational convention for Weyl reflections of length 2 \ ( $\ell = 2$ ), is
$$ w_{i_1,i_2}(\mu) \equiv w_{i_1}(w_{i_2}(\mu))
\eqno(II.4) $$
while for length k \ ($\ell = k$ ), its generalization is
$$ w_{i_1,i_2,.. ,i_k}(\mu) \equiv w_{i_1}(w_{i_2}(.. w_{i_k}(\mu)).. )
\eqno(II.5) $$
on any weight $\mu$. The Weyl group $W(E_7)$ is then a finite group which
is constructed out of a finite number of products of these simple reflections.
It is also known that $W(E_7)$ contains a unity of length zero and also
an element of maximum length which is equal to the number of positive
roots of $E_7$. To be in a complete compatibility with the section II of
ref.[1], one can readily seen that ,when they are applied on a strictly
dominant weight $\Lambda^{++}$ of $E_7$, the following elements of $W(E_7)$
give us 36 of elements of the set $\varphi(\Lambda^{++})$ of permutation
weights:
$$ \vbox{\tabskip=0pt \offinterlineskip
\halign to420pt{\strut#& \vrule#\tabskip=1em plus2em& \hfil#& \vrule#&
\hfil#\hfil& \vrule#\tabskip=0pt\cr \noalign{\hrule}
& & $\ell=0$ & & $\eqalign{&~ \cr &1 \cr &~}$ & \cr  \noalign{\hrule}
& & $\ell=1$ & & $\eqalign{&~ \cr &w_3 \cr &~}$ & \cr  \noalign{\hrule}
& & $\ell=2$ & & $\eqalign{&~ \cr &w_{3,2}, w_{3,4} \cr &~}$ & \cr \noalign{\hrule}
& & $\ell=3$ & &
$\eqalign{&~ \cr &w_{3,2,4} \ , \ w_{3,4,5} \ , \ w_{3,4,7} \cr &~}$ & \cr \noalign{\hrule}
& & $\ell=4$  & &
$\eqalign{&~ \cr &w_{3,2,4,3} \ , \
           w_{3,2,4,5} \ , \
           w_{3,2,4,7} \ , \
           w_{3,4,5,6} \ , \
           w_{3,4,5,7} \cr &~}$ & \cr  \noalign{\hrule}
& & $\ell=5$  & &
$\eqalign{&~ \cr &w_{3,2,4,3,5} \ , \
           w_{3,2,4,3,7} \ , \
           w_{3,2,4,5,6} \ , \
           w_{3,2,4,5,7} \ , \
           w_{3,4,5,6,7} \ , \
           w_{3,4,5,7,4} \cr &~}$  & \cr  \noalign{\hrule}
& & $\ell=6$  & &
$\eqalign{&~ \cr &w_{3,2,4,3,5,4} \ , \
           w_{3,2,4,3,5,6} \ , \
           w_{3,2,4,3,5,7} \ , \
           w_{3,2,4,3,7,4} \cr
          &w_{3,2,4,5,6,7} \ , \
           w_{3,2,4,5,7,4} \ , \
           w_{3,4,5,6,7,4} \ , \
           w_{3,4,5,7,4,3} \cr &~}$ & \cr  \noalign{\hrule}
& & $\ell=7$  & &
$\eqalign{&~ \cr &w_{3,2,4,3,5,4,6} \ , \
           w_{3,2,4,3,5,6,7} \ , \
           w_{3,2,4,5,6,7,4} \ , \
           w_{3,4,5,6,7,4,3} \ , \
           w_{3,4,5,6,7,4,5} \cr &~}$ & \cr  \noalign{\hrule}
& & $\ell=8$  & &
$\eqalign{&~ \cr &w_{3,2,4,3,5,4,6,5} \ , \
           w_{3,2,4,5,6,7,4,5} \ , \
           w_{3,4,5,6,7,4,3,5} \cr &~}$ & \cr  \noalign{\hrule}
& & $\ell=9$  & &
$\eqalign{&~ \cr &w_{3,4,5,6,7,4,3,5,4} \cr &~}$ & \cr  \noalign{\hrule}
& & $\ell=10$  & &
$\eqalign{&~ \cr &w_{3,4,5,6,7,4,3,5,4,7} \cr &~}$ & \cr  \noalign{\hrule}
}} $$
\centerline{Table-1}
\vskip4mm
\noindent For any one of these Weyl reflections $w^{(K)}$ \ (K=1,2,.. ,36),
corresponding signature $\epsilon(K) \ \ ( \ \equiv \epsilon(w^{K}) \ )$
is known to be assigned by
$$ \epsilon(K) = (-1)^{\ell(w^{K})}   \eqno(II.7) $$
\noindent Remaining 36 elements of $\varphi(\Lambda^{++})$ can be seen to be
obtained by applying $A_7$ diagram automorhism but with opposite signatures.

As is also shown in ref.[1], all these solve in principle the problem of
summing over $E_7$ Weyl group by reducing the problem over the Weyl group
of $A_7$. In the last section, we will show however that there are still
more to made for bringing a completely practical solution to this problem.

\vskip 3mm
\noindent {\bf III.\ ${\bf (E_8 , A_8)}$ }
\vskip 3mm

The procedure here is completely similar to the one given in section II.
Signatures of a Weyl reflection will be specified as in (II.7) though
one important notice here is that, contrary to the case $E_7$, signatures
remain unchanged under $A_8$ diagram automorphism. Similarly as in (II.1),
the decompositions of $E_8$ Weyl orbits in terms of $A_8$ Weyl orbits have
been given elsewhere {\bf [4]}. A way to reduce $E_8$ into $A_8$ is then as in
the following:
$$ \eqalign{
\Lambda_1 &= \lambda_3                                          \cr
\Lambda_2 &= \lambda_3+\lambda_6                                \cr
\Lambda_3 &= \lambda_3+\lambda_5+\lambda_7                      \cr
\Lambda_4 &= \lambda_3+\lambda_5+\lambda_5+\lambda_8            \cr
\Lambda_5 &= \lambda_2+\lambda_4+\lambda_5+\lambda_5+\lambda_8  \cr
\Lambda_6 &= \lambda_1+\lambda_4+\lambda_5+\lambda_8            \cr
\Lambda_7 &= \lambda_4+\lambda_8                                \cr
\Lambda_8 &= \lambda_4+\lambda_5              } \eqno(III.1) $$
\noindent For any strictly dominant weight $\Lambda^{++}$ of $E_8$, 1920
elements of $\varphi(\Lambda^{++})$ will be obtained when 1920 appropriately
chosen elements of $W(E_8)$ act on $\Lambda^{++}$. First 960 elements of
$W(E_8)$ are given in the appendix. Remaining 960 elements can be obtained via
$A_8$ diagram automorphism given by $ \lambda_i \rightarrow \lambda_{9-i}$.

\vskip 3mm
\noindent {\bf IV.\ ${\bf (A_8 , A_7)}$ }
\vskip 3mm

As is frequently emphasized above, the problem of making summations over
Weyl groups of $E_7$ and also $E_8$ has now been reduced over the Weyl groups
of $A_7$ and $A_8$ within the framework of the method proposed in sections II
and III. For an experienced reader, this is still quite problematic in the
absence of an extra simplification concerning  pairs $(A_N , A_{N-1})$.
In fact, this has been already given in an unpublished work {\bf [5]}.
We find however convenient to consider the problem something more closely here.

To be more concrete, let us proceed in the example $(A_8 , A_7)$ which has
a central role in this work. The generalization to any other pair
$(A_N , A_{N-1})$ will however be trivial. Let $\Lambda_A$ \ ($\lambda_a$) be
the fundamental dominant weights of $A_8$ \ ($A_7$) for A=1,2,.. 8 \
(a=1,2,..7). The nine fundamental weights $\mu_I$ of $A_8$  are defined,
for I=1,2,...9 , as in section I. As any other dominant weight, a strictly
dominant weight $\Lambda^{++}$ is expressed in the form
$$ \Lambda^{++} = \sum_{A=1}^8 \ i_A \ \mu_A   \eqno(IV.1)  $$
on condition that $i_A$'s are some positive integers constrained by
$$ i_1 > i_2 > ... >i_8  \ \ . \eqno(IV.2) $$
We now know as a result of $dimW(A_8)/dimW(A_7) = 9$ that
$\varphi(\Lambda^{++})$ has always 9 elements. If one rewrites (IV.1)
in the trivial notation of
$$ \Lambda^{++} = ( \ i_1 , i_2 , i_3 , i_4 , i_5 , i_6 , i_7 , i_8 , 0 )
\eqno(IV.3)  $$
following expressions will be obtained for these 9 elements $\Lambda^{++}(I)$:
$$ \eqalign{
\Lambda^{++}(1) &= ( \ i_1 , i_2 , i_3 , i_4 , i_5 , i_6 , i_7 , i_8 , 0 \ )  \cr
\Lambda^{++}(2) &= ( \ i_1 , i_2 , i_3 , i_4 , i_5 , i_6 , i_7 , 0 , i_8 \ )  \cr
\Lambda^{++}(3) &= ( \ i_1 , i_2 , i_3 , i_4 , i_5 , i_6 , i_8 , 0 , i_7 \ )  \cr
\Lambda^{++}(4) &= ( \ i_1 , i_2 , i_3 , i_4 , i_5 , i_7 , i_8 , 0 , i_6 \ )  \cr
\Lambda^{++}(5) &= ( \ i_1 , i_2 , i_3 , i_4 , i_6 , i_7 , i_8 , 0 , i_5 \ )  \cr
\Lambda^{++}(6) &= ( \ i_1 , i_2 , i_3 , i_5 , i_6 , i_7 , i_8 , 0 , i_4 \ )  \cr
\Lambda^{++}(7) &= ( \ i_1 , i_2 , i_4 , i_5 , i_6 , i_7 , i_8 , 0 , i_3 \ )  \cr
\Lambda^{++}(8) &= ( \ i_1 , i_3 , i_4 , i_5 , i_6 , i_7 , i_8 , 0 , i_2 \ )  \cr
\Lambda^{++}(9) &= ( \ i_2 , i_3 , i_4 , i_5 , i_6 , i_7 , i_8 , 0 , i_1 \ )  } \eqno(IV.4) $$
\noindent Signatures which enter in summations over $W(A_8)$ can be assigned by
$$ \epsilon(\Lambda^{++}(I)) = (-1)^{I+1} \ , \ I=1,2,...9.    \eqno(IV.5) $$
The logic here is that (IV.5) is to be obtained by enumerating transpositions
which reduce any $\Lambda^{++}(I)$ to $\Lambda^{++}(1)$ having
$\epsilon(\Lambda^{++}(1))$ = +1 by definition.

The central point here is the fact that only $\Lambda^{++}(1)$ is itself an
$A_8$ dominant weight though first 8 components of all the 9 weights in (IV.4)
provide us $A_7$ dominant weights as will be seen by the aid of the following
reduction of $A_8$ to $A_7$:
$$ \Lambda_a = \lambda_a + {a \over 8} \ \Lambda_8  \ , \ a=1,2,..7. \eqno(IV.6) $$
It can now be seen that the specialization
$$ e^{\mu_A} \equiv u_A \ \ , \ \ e^{\Lambda_8} \equiv 1 \ \ , \ \
A=1,2,..8.   \eqno(IV.7) $$
of formal exponentials {\bf [6]} casts the $A_8$ problem completely into a
$A_7$ problem which is subject to the contraint
$$ u_1 u_2 ... u_8 \equiv 1 \ \ . \eqno(IV.8) $$
The degenerated Schur functions which are introduced and shown to be the
subjects of some reduction formulas given in an unpublished work {\bf [5]},
must now be determined by the aid of parameters $u_A$ for $A_7$ Lie algebra.
To be more instructive, let us consider
$$ A(\Lambda^{++}) \equiv \sum_{\mu \in W(\Lambda^{++})} \epsilon(\mu) \
e^\mu $$
and remark that all these can be summarized by the identification that
$$ A(\Lambda^{++}) \equiv \sum_{I=1}^9 A(\lambda^{++}(I)) \eqno(IV.9) $$
\noindent where $\lambda^{++}(I)$'s are $A_7$ dominant weights which are
specified by (IV.4) as in the following:
$$ \eqalign{
\lambda^{++}(1) &= ( \ i_1 , i_2 , i_3 , i_4 , i_5 , i_6 , i_7 , i_8 \ )  \cr
\lambda^{++}(2) &= ( \ i_1 , i_2 , i_3 , i_4 , i_5 , i_6 , i_7 ,  0  \ )  \cr
\lambda^{++}(3) &= ( \ i_1 , i_2 , i_3 , i_4 , i_5 , i_6 , i_8 ,  0  \ )  \cr
\lambda^{++}(4) &= ( \ i_1 , i_2 , i_3 , i_4 , i_5 , i_7 , i_8 ,  0  \ )  \cr
\lambda^{++}(5) &= ( \ i_1 , i_2 , i_3 , i_4 , i_6 , i_7 , i_8 ,  0  \ )  \cr
\lambda^{++}(6) &= ( \ i_1 , i_2 , i_3 , i_5 , i_6 , i_7 , i_8 ,  0  \ )  \cr
\lambda^{++}(7) &= ( \ i_1 , i_2 , i_4 , i_5 , i_6 , i_7 , i_8 ,  0  \ )  \cr
\lambda^{++}(8) &= ( \ i_1 , i_3 , i_4 , i_5 , i_6 , i_7 , i_8 ,  0  \ )  \cr
\lambda^{++}(9) &= ( \ i_2 , i_3 , i_4 , i_5 , i_6 , i_7 , i_8 ,  0  \ )  } \eqno(IV.10) $$

Careful reader will notice that above procedure allows us to reduce an $A_N$
problem down to some quite low value of the rank N. For simple applications
of Mathematicae {\bf[7]}, $A_5$ will be sufficiently plausible to get
accoustomed with an $E_8$ or $E_7$ calculation.

\vskip3mm
\noindent{\bf REFERENCES}
\vskip3mm

\leftline{[1] H.R.Karadayi and M.Gungormez, Fundamental Weights, Permutations Weights and }
\leftline{ \ \ \ \ Weyl Character Formula, to appear in Journal of Physics A: Mathematical and General  }

\leftline{[2] J.E.Humphreys, Introduction to Lie Algebras and Representation Theory, N.Y., Springer-Verlag (1972)}

\leftline{[3] H.R.Karadayi and M.Gungormez, Jour.Math.Phys., 38 (1997) 5991-6007 }
\leftline{ \ \ \ \ H.R.Karadayi, Anatomy of Grand Unifying Groups I and II ,}
\leftline{ \ \ \ \ ICTP preprints(unpublished) IC/81/213 and 224 }

\leftline{[4] H.R.Karadayi and M.Gungormez, The Higher Cohomologies of $E_8$ Lie Algebra, physics/9701004 }

\leftline{[5] H.R.Karadayi, $A_N$ Multiplicity Rules and Schur Functions (unpublished) , math-ph/9805009  }

\leftline{[6] V.G.Kac, Infinite Dimensional Lie Algebras, N.Y., Cambridge Univ. Press (1990)}

\leftline{[7] S. Wolfram, Mathematica$^{TM}$, Addison-Wesley (1990) }

\vskip5mm
\vskip5mm

\noindent{\bf APPENDIX  }

$$ \vbox{\tabskip=0pt \offinterlineskip
\halign to420pt{\strut#& \vrule#\tabskip=1em plus2em& \hfil#& \vrule#&
\hfil#\hfil& \vrule#\tabskip=0pt\cr \noalign{\hrule}
& & $\ell=0$ & & $\eqalign{&~ \cr &1 \cr &~}$ & \cr  \noalign{\hrule}
& & $\ell=1$ & & $\eqalign{&~ \cr
      &w_1 \ , \
       w_4 \ , \
       w_7 \ , \
       w_8 \cr &~}$ & \cr  \noalign{\hrule}
& & $\ell=2$ & & $\eqalign{&~ \cr
      &w_{1,2} \ , \
       w_{1,4} \ , \
       w_{1,7} \ , \
       w_{1,8} \ , \
       w_{4,3} \ , \
       w_{4,5} \cr
       &w_{4,7} \ , \
       w_{4,8} \ , \
       w_{7,6} \ , \
       w_{7,8} \ , \
       w_{8,5} \cr &~}$ & \cr \noalign{\hrule}
& & $\ell=3$ & &
$\eqalign{&~ \cr
     &w_{1,2,3} \ , \
      w_{1,2,4} \ , \
      w_{1,2,7} \ , \
      w_{1,2,8} \ , \
      w_{1,4,3} \ , \
      w_{1,4,5} \cr
     &w_{1,4,7} \ , \
      w_{1,4,8} \ , \
      w_{1,7,6} \ , \
      w_{1,7,8} \ , \
      w_{1,8,5} \ , \
      w_{4,3,2} \cr
     &w_{4,3,8} \ , \
      w_{4,5,6} \ , \
      w_{4,5,8} \ , \
      w_{4,7,6} \ , \
      w_{4,7,8} \ , \
      w_{4,8,5} \cr
     &w_{7,6,5} \ , \
      w_{7,6,8} \ , \
      w_{7,8,5} \ , \
      w_{8,5,4} \ , \
      w_{8,5,6} \cr &~}$ & \cr \noalign{\hrule}
& & $\ell=4$  & &
$\eqalign{&~ \cr
     &w_{1,2,3,7} \ , \
      w_{1,2,3,8} \ , \
      w_{1,2,4,5} \ , \
      w_{1,2,4,7} \ , \
      w_{1,2,4,8} \ , \
      w_{1,2,7,6} \cr
      &w_{1,2,7,8} \ , \
      w_{1,2,8,5} \ , \
      w_{1,4,3,8} \ , \
      w_{1,4,5,6} \ , \
      w_{1,4,5,8} \ , \
      w_{1,4,7,6} \cr
      &w_{1,4,7,8} \ , \
      w_{1,4,8,5} \ , \
      w_{1,7,6,5} \ , \
      w_{1,7,6,8} \ , \
      w_{1,7,8,5} \ , \
      w_{1,8,5,4} \cr
      &w_{1,8,5,6} \ , \
      w_{4,3,2,8} \ , \
      w_{4,3,8,5} \ , \
      w_{4,5,6,8} \ , \
      w_{4,5,8,5} \ , \
      w_{4,7,6,8} \cr
      &w_{4,7,8,5} \ , \
      w_{4,8,5,4} \ , \
      w_{4,8,5,6} \ , \
      w_{7,8,5,4} \ , \
      w_{8,5,4,3} \ , \
      w_{8,5,4,6} \cr &~}$ & \cr  \noalign{\hrule}  }} $$
\centerline{Table-2}

$$ \vbox{\tabskip=0pt \offinterlineskip
\halign to420pt{\strut#& \vrule#\tabskip=1em plus2em& \hfil#& \vrule#&
\hfil#\hfil& \vrule#\tabskip=0pt\cr \noalign{\hrule}
& & $\ell=5$  & &
$\eqalign{&~ \cr
     &w_{1,2,3,7,6} \ , \
      w_{1,2,3,7,8} \ , \
      w_{1,2,3,8,5} \ , \
      w_{1,2,4,5,6} \ , \
      w_{1,2,4,5,8} \ , \
      w_{1,2,4,7,6} \ , \
      w_{1,2,4,7,8} \cr
     &w_{1,2,4,8,5} \ , \
      w_{1,2,7,6,5} \ , \
      w_{1,2,7,6,8} \ , \
      w_{1,2,7,8,5} \ , \
      w_{1,2,8,5,4} \ , \
      w_{1,2,8,5,6} \ , \
      w_{1,4,3,8,5} \cr
     &w_{1,4,5,6,8} \ , \
      w_{1,4,5,8,5} \ , \
      w_{1,4,7,6,8} \ , \
      w_{1,4,7,8,5} \ , \
      w_{1,4,8,5,4} \ , \
      w_{1,4,8,5,6} \ , \
      w_{1,7,8,5,4} \cr
     &w_{1,8,5,4,3} \ , \
      w_{1,8,5,4,6} \ , \
      w_{4,3,2,8,5} \ , \
      w_{4,3,8,5,4} \ , \
      w_{4,3,8,5,6} \ , \
      w_{4,5,6,8,5} \ , \
      w_{4,5,8,5,4} \cr
     &w_{4,5,8,5,6} \ , \
      w_{4,7,6,8,5} \ , \
      w_{4,7,8,5,4} \ , \
      w_{4,7,8,5,6} \ , \
      w_{4,8,5,4,3} \ , \
      w_{4,8,5,4,6} \ , \
      w_{4,8,5,6,7} \cr
     &w_{7,8,5,4,3} \ , \
      w_{8,5,4,3,2} \ , \
      w_{8,5,4,3,6} \ , \
      w_{8,5,4,6,5} \cr &~}$ & \cr  \noalign{\hrule}
& & $\ell=6$  & &
$\eqalign{&~ \cr
     &w_{1,2,3,7,6,5} \ , \
      w_{1,2,3,7,6,8} \ , \
      w_{1,2,3,7,8,5} \ , \
      w_{1,2,3,8,5,4} \ , \
      w_{1,2,3,8,5,6} \ , \
      w_{1,2,4,5,6,8} \cr
     &w_{1,2,4,5,8,5} \ , \
      w_{1,2,4,7,6,8} \ , \
      w_{1,2,4,7,8,5} \ , \
      w_{1,2,4,8,5,4} \ , \
      w_{1,2,4,8,5,6} \ , \
      w_{1,2,7,8,5,4} \cr
     &w_{1,2,8,5,4,3} \ , \
      w_{1,2,8,5,4,6} \ , \
      w_{1,4,3,8,5,4} \ , \
      w_{1,4,3,8,5,6} \ , \
      w_{1,4,5,6,8,5} \ , \
      w_{1,4,5,8,5,4} \cr
     &w_{1,4,5,8,5,6} \ , \
      w_{1,4,7,6,8,5} \ , \
      w_{1,4,7,8,5,4} \ , \
      w_{1,4,7,8,5,6} \ , \
      w_{1,4,8,5,4,3} \ , \
      w_{1,4,8,5,4,6} \cr
     &w_{1,4,8,5,6,7} \ , \
      w_{1,7,8,5,4,3} \ , \
      w_{1,8,5,4,3,2} \ , \
      w_{1,8,5,4,3,6} \ , \
      w_{1,8,5,4,6,5} \ , \
      w_{4,3,2,8,5,4} \cr
     &w_{4,3,2,8,5,6} \ , \
      w_{4,3,8,5,4,3} \ , \
      w_{4,3,8,5,4,6} \ , \
      w_{4,3,8,5,6,7} \ , \
      w_{4,5,6,8,5,4} \ , \
      w_{4,5,6,8,5,6} \cr
     &w_{4,5,8,5,4,3} \ , \
      w_{4,5,8,5,4,6} \ , \
      w_{4,5,8,5,6,7} \ , \
      w_{4,7,6,8,5,6} \ , \
      w_{4,7,8,5,4,3} \ , \
      w_{4,7,8,5,4,6} \cr
     &w_{4,7,8,5,6,7} \ , \
      w_{4,8,5,4,3,2} \ , \
      w_{4,8,5,4,3,6} \ , \
      w_{4,8,5,4,6,5} \ , \
      w_{4,8,5,4,6,7} \ , \
      w_{7,8,5,4,3,2} \cr
     &w_{8,5,4,3,2,1} \ , \
      w_{8,5,4,3,2,6} \ , \
      w_{8,5,4,3,6,5} \ , \
      w_{8,5,4,6,5,8} \cr  &~}$ & \cr  \noalign{\hrule}
& & $\ell=7$  & &
$\eqalign{&~ \cr
     &w_{1,2,3,7,6,5,4}  \ , \
      w_{1,2,3,7,8,5,4}  \ , \
      w_{1,2,3,8,5,4,3}  \ , \
      w_{1,2,3,8,5,4,6}  \ , \
      w_{1,2,4,5,6,8,5}  \cr
     &w_{1,2,4,5,8,5,4}  \ , \
      w_{1,2,4,5,8,5,6}  \ , \
      w_{1,2,4,7,6,8,5}  \ , \
      w_{1,2,4,7,8,5,4}  \ , \
      w_{1,2,4,7,8,5,6}  \cr
     &w_{1,2,4,8,5,4,3}  \ , \
      w_{1,2,4,8,5,4,6}  \ , \
      w_{1,2,4,8,5,6,7}  \ , \
      w_{1,2,7,8,5,4,3}  \ , \
      w_{1,2,8,5,4,3,2}  \cr
     &w_{1,2,8,5,4,3,6}  \ , \
      w_{1,2,8,5,4,6,5}  \ , \
      w_{1,4,3,8,5,4,3}  \ , \
      w_{1,4,3,8,5,4,6}  \ , \
      w_{1,4,3,8,5,6,7}  \cr
     &w_{1,4,5,6,8,5,4}  \ , \
      w_{1,4,5,6,8,5,6}  \ , \
      w_{1,4,5,8,5,4,3}  \ , \
      w_{1,4,5,8,5,4,6}  \ , \
      w_{1,4,5,8,5,6,7}  \cr
     &w_{1,4,7,6,8,5,6}  \ , \
      w_{1,4,7,8,5,4,3}  \ , \
      w_{1,4,7,8,5,4,6}  \ , \
      w_{1,4,7,8,5,6,7}  \ , \
      w_{1,4,8,5,4,3,2}  \cr
     &w_{1,4,8,5,4,3,6}  \ , \
      w_{1,4,8,5,4,6,5}  \ , \
      w_{1,4,8,5,4,6,7}  \ , \
      w_{1,7,8,5,4,3,2}  \ , \
      w_{1,8,5,4,3,2,1}  \cr
     &w_{1,8,5,4,3,2,6}  \ , \
      w_{1,8,5,4,3,6,5}  \ , \
      w_{1,8,5,4,6,5,8}  \ , \
      w_{4,3,2,8,5,4,3}  \ , \
      w_{4,3,2,8,5,4,6}  \cr
     &w_{4,3,2,8,5,6,7}  \ , \
      w_{4,3,8,5,4,3,2}  \ , \
      w_{4,3,8,5,4,3,6}  \ , \
      w_{4,3,8,5,4,6,5}  \ , \
      w_{4,3,8,5,4,6,7}  \cr
     &w_{4,5,6,8,5,4,6}  \ , \
      w_{4,5,6,8,5,6,7}  \ , \
      w_{4,5,8,5,4,3,2}  \ , \
      w_{4,5,8,5,4,3,6}  \ , \
      w_{4,5,8,5,4,6,5}  \cr
     &w_{4,5,8,5,4,6,7}  \ , \
      w_{4,7,6,8,5,6,7}  \ , \
      w_{4,7,8,5,4,3,2}  \ , \
      w_{4,7,8,5,4,3,6}  \ , \
      w_{4,7,8,5,4,6,7}  \cr
     &w_{4,8,5,4,3,2,1}  \ , \
      w_{4,8,5,4,3,2,6}  \ , \
      w_{4,8,5,4,3,6,5}  \ , \
      w_{4,8,5,4,3,6,7}  \ , \
      w_{4,8,5,4,6,5,7}  \cr
     &w_{4,8,5,4,6,5,8}  \ , \
      w_{7,8,5,4,3,2,1}  \ , \
      w_{8,5,4,3,2,1,6}  \ , \
      w_{8,5,4,3,2,6,5}  \ , \
      w_{8,5,4,3,6,5,4}  \cr
     &w_{8,5,4,3,6,5,8}  \cr  &~}$ & \cr  \noalign{\hrule}
& & $\ell=8$  & &
$\eqalign{&~ \cr
	  &w_{1,2,3,7,8,5,4,3} \ , \
	   w_{1,2,3,8,5,4,3,2} \ , \
	   w_{1,2,3,8,5,4,3,6} \ , \
	   w_{1,2,3,8,5,4,6,5} \ , \
	   w_{1,2,4,5,6,8,5,4} \cr
	  &w_{1,2,4,5,6,8,5,6} \ , \
	   w_{1,2,4,5,8,5,4,3} \ , \
	   w_{1,2,4,5,8,5,4,6} \ , \
	   w_{1,2,4,5,8,5,6,7} \ , \
	   w_{1,2,4,7,6,8,5,6} \cr
	  &w_{1,2,4,7,8,5,4,3} \ , \
	   w_{1,2,4,7,8,5,4,6} \ , \
	   w_{1,2,4,7,8,5,6,7} \ , \
	   w_{1,2,4,8,5,4,3,2} \ , \
	   w_{1,2,4,8,5,4,3,6} \cr
	  &w_{1,2,4,8,5,4,6,5} \ , \
	   w_{1,2,4,8,5,4,6,7} \ , \
	   w_{1,2,7,8,5,4,3,2} \ , \
	   w_{1,2,8,5,4,3,2,1} \ , \
	   w_{1,2,8,5,4,3,2,6} \cr  &~}$ & \cr  \noalign{\hrule}
}}$$

$$ \vbox{\tabskip=0pt \offinterlineskip
\halign to420pt{\strut#& \vrule#\tabskip=1em plus2em& \hfil#& \vrule#&
\hfil#\hfil& \vrule#\tabskip=0pt\cr \noalign{\hrule}
& & & &
$\eqalign{&~ \cr
	  &w_{1,2,8,5,4,3,6,5} \ , \
	   w_{1,2,8,5,4,6,5,8} \ , \
	   w_{1,4,3,8,5,4,3,2} \ , \
	   w_{1,4,3,8,5,4,3,6} \ , \
	   w_{1,4,3,8,5,4,6,5} \cr
	  &w_{1,4,3,8,5,4,6,7} \ , \
	   w_{1,4,5,6,8,5,4,6} \ , \
	   w_{1,4,5,6,8,5,6,7} \ , \
	   w_{1,4,5,8,5,4,3,2} \ , \
	   w_{1,4,5,8,5,4,3,6} \cr
	  &w_{1,4,5,8,5,4,6,5} \ , \
	   w_{1,4,5,8,5,4,6,7} \ , \
	   w_{1,4,7,6,8,5,6,7} \ , \
	   w_{1,4,7,8,5,4,3,2} \ , \
	   w_{1,4,7,8,5,4,3,6} \cr
	  &w_{1,4,7,8,5,4,6,7} \ , \
	   w_{1,4,8,5,4,3,2,1} \ , \
	   w_{1,4,8,5,4,3,2,6} \ , \
	   w_{1,4,8,5,4,3,6,5} \ , \
	   w_{1,4,8,5,4,3,6,7} \cr
	  &w_{1,4,8,5,4,6,5,7} \ , \
	   w_{1,4,8,5,4,6,5,8} \ , \
	   w_{1,7,8,5,4,3,2,1} \ , \
	   w_{1,8,5,4,3,2,1,6} \ , \
	   w_{1,8,5,4,3,2,6,5} \cr
	  &w_{1,8,5,4,3,6,5,4} \ , \
	   w_{1,8,5,4,3,6,5,8} \ , \
	   w_{4,3,2,8,5,4,3,2} \ , \
	   w_{4,3,2,8,5,4,3,6} \ , \
	   w_{4,3,2,8,5,4,6,5} \cr
	  &w_{4,3,2,8,5,4,6,7} \ , \
	   w_{4,3,8,5,4,3,2,1} \ , \
	   w_{4,3,8,5,4,3,2,6} \ , \
	   w_{4,3,8,5,4,3,6,5} \ , \
	   w_{4,3,8,5,4,3,6,7} \cr
	  &w_{4,3,8,5,4,6,5,7} \ , \
	   w_{4,5,6,8,5,4,6,5} \ , \
	   w_{4,5,6,8,5,4,6,7} \ , \
	   w_{4,5,8,5,4,3,2,1} \ , \
	   w_{4,5,8,5,4,3,2,6} \cr
	  &w_{4,5,8,5,4,3,6,5} \ , \
	   w_{4,5,8,5,4,3,6,7} \ , \
	   w_{4,5,8,5,4,6,5,7} \ , \
	   w_{4,5,8,5,4,6,5,8} \ , \
	   w_{4,7,8,5,4,3,2,1} \cr
	  &w_{4,7,8,5,4,3,2,6} \ , \
	   w_{4,7,8,5,4,3,6,7} \ , \
	   w_{4,8,5,4,3,2,1,6} \ , \
	   w_{4,8,5,4,3,2,6,5} \ , \
	   w_{4,8,5,4,3,2,6,7} \cr
	  &w_{4,8,5,4,3,6,5,4} \ , \
	   w_{4,8,5,4,3,6,5,7} \ , \
	   w_{4,8,5,4,3,6,5,8} \ , \
	   w_{4,8,5,4,6,5,7,8} \ , \
	   w_{8,5,4,3,2,1,6,5} \cr
	  &w_{8,5,4,3,2,6,5,4} \ , \
	   w_{8,5,4,3,2,6,5,8} \ , \
	   w_{8,5,4,3,6,5,4,8} \cr  &~}$ & \cr  \noalign{\hrule}
& & $\ell=9$  & &
$\eqalign{&~ \cr
	 &w_{1,2,3,7,8,5,4,3,2}  \ , \
	  w_{1,2,3,8,5,4,3,2,1}  \ , \
	  w_{1,2,3,8,5,4,3,2,6}  \ , \
	  w_{1,2,3,8,5,4,3,6,5}  \cr
	 &w_{1,2,3,8,5,4,6,5,8}  \ , \
	  w_{1,2,4,5,6,8,5,4,3}  \ , \
	  w_{1,2,4,5,6,8,5,4,6}  \ , \
	  w_{1,2,4,5,6,8,5,6,7}  \cr
	 &w_{1,2,4,5,8,5,4,3,2}  \ , \
	  w_{1,2,4,5,8,5,4,3,6}  \ , \
	  w_{1,2,4,5,8,5,4,6,5}  \ , \
	  w_{1,2,4,5,8,5,4,6,7}  \cr
	 &w_{1,2,4,7,6,8,5,6,7}  \ , \
	  w_{1,2,4,7,8,5,4,3,2}  \ , \
	  w_{1,2,4,7,8,5,4,3,6}  \ , \
	  w_{1,2,4,7,8,5,4,6,7}  \cr
	 &w_{1,2,4,8,5,4,3,2,1}  \ , \
	  w_{1,2,4,8,5,4,3,2,6}  \ , \
	  w_{1,2,4,8,5,4,3,6,5}  \ , \
	  w_{1,2,4,8,5,4,3,6,7}  \cr
	 &w_{1,2,4,8,5,4,6,5,7}  \ , \
	  w_{1,2,4,8,5,4,6,5,8}  \ , \
	  w_{1,2,7,8,5,4,3,2,1}  \ , \
	  w_{1,2,8,5,4,3,2,1,6}  \cr
	 &w_{1,2,8,5,4,3,2,6,5}  \ , \
	  w_{1,2,8,5,4,3,6,5,4}  \ , \
	  w_{1,2,8,5,4,3,6,5,8}  \ , \
	  w_{1,4,3,8,5,4,3,2,1}  \cr
	 &w_{1,4,3,8,5,4,3,2,6}  \ , \
	  w_{1,4,3,8,5,4,3,6,5}  \ , \
	  w_{1,4,3,8,5,4,3,6,7}  \ , \
	  w_{1,4,3,8,5,4,6,5,7}  \cr
	 &w_{1,4,5,6,8,5,4,6,5}  \ , \
	  w_{1,4,5,6,8,5,4,6,7}  \ , \
	  w_{1,4,5,8,5,4,3,2,1}  \ , \
	  w_{1,4,5,8,5,4,3,2,6}  \cr
	 &w_{1,4,5,8,5,4,3,6,5}  \ , \
	  w_{1,4,5,8,5,4,3,6,7}  \ , \
	  w_{1,4,5,8,5,4,6,5,7}  \ , \
	  w_{1,4,5,8,5,4,6,5,8}  \cr
	 &w_{1,4,7,8,5,4,3,2,1}  \ , \
	  w_{1,4,7,8,5,4,3,2,6}  \ , \
	  w_{1,4,7,8,5,4,3,6,7}  \ , \
	  w_{1,4,8,5,4,3,2,1,6}  \cr
	 &w_{1,4,8,5,4,3,2,6,5}  \ , \
	  w_{1,4,8,5,4,3,2,6,7}  \ , \
	  w_{1,4,8,5,4,3,6,5,4}  \ , \
	  w_{1,4,8,5,4,3,6,5,7}  \cr
	 &w_{1,4,8,5,4,3,6,5,8}  \ , \
	  w_{1,4,8,5,4,6,5,7,8}  \ , \
	  w_{1,8,5,4,3,2,1,6,5}  \ , \
	  w_{1,8,5,4,3,2,6,5,4}  \cr
	 &w_{1,8,5,4,3,2,6,5,8}  \ , \
	  w_{1,8,5,4,3,6,5,4,8}  \ , \
	  w_{4,3,2,8,5,4,3,2,1}  \ , \
	  w_{4,3,2,8,5,4,3,2,6}  \cr
	 &w_{4,3,2,8,5,4,3,6,5}  \ , \
	  w_{4,3,2,8,5,4,3,6,7}  \ , \
	  w_{4,3,2,8,5,4,6,5,7}  \ , \
	  w_{4,3,8,5,4,3,2,1,6}  \cr
	 &w_{4,3,8,5,4,3,2,6,5}  \ , \
	  w_{4,3,8,5,4,3,2,6,7}  \ , \
	  w_{4,3,8,5,4,3,6,5,4}  \ , \
	  w_{4,3,8,5,4,3,6,5,7}  \cr
	 &w_{4,3,8,5,4,6,5,7,6}  \ , \
	  w_{4,5,6,8,5,4,6,5,7}  \ , \
	  w_{4,5,6,8,5,4,6,5,8}  \ , \
	  w_{4,5,8,5,4,3,2,1,6}  \cr
	 &w_{4,5,8,5,4,3,2,6,5}  \ , \
	  w_{4,5,8,5,4,3,2,6,7}  \ , \
	  w_{4,5,8,5,4,3,6,5,7}  \ , \
	  w_{4,5,8,5,4,3,6,5,8}  \cr
	 &w_{4,5,8,5,4,6,5,7,6}  \ , \
	  w_{4,5,8,5,4,6,5,7,8}  \ , \
	  w_{4,7,8,5,4,3,2,1,6}  \ , \
	  w_{4,7,8,5,4,3,2,6,7}  \cr
	 &w_{4,8,5,4,3,2,1,6,5}  \ , \
	  w_{4,8,5,4,3,2,1,6,7}  \ , \
	  w_{4,8,5,4,3,2,6,5,4}  \ , \
	  w_{4,8,5,4,3,2,6,5,7}  \cr
	 &w_{4,8,5,4,3,2,6,5,8}  \ , \
	  w_{4,8,5,4,3,6,5,4,7}  \ , \
	  w_{4,8,5,4,3,6,5,4,8}  \ , \
	  w_{4,8,5,4,3,6,5,7,8}  \cr
	 &w_{8,5,4,3,2,1,6,5,4}  \ , \
	  w_{8,5,4,3,2,1,6,5,8}  \ , \
	  w_{8,5,4,3,2,6,5,4,3}  \ , \
	  w_{8,5,4,3,2,6,5,4,8}  \cr
	 &w_{8,5,4,3,6,5,4,8,5}  \cr  &~}$ & \cr  \noalign{\hrule}
}}$$

$$ \vbox{\tabskip=0pt \offinterlineskip
\halign to450pt{\strut#& \vrule#\tabskip=1em plus2em& \hfil#& \vrule#&
\hfil#\hfil& \vrule#\tabskip=0pt\cr \noalign{\hrule}
& & $\ell=10$ & &
$\eqalign{&~ \cr
	 &w_{1,2,3,7,8,5,4,3,2,1} \ , \
	  w_{1,2,3,8,5,4,3,2,1,6} \ , \
	  w_{1,2,3,8,5,4,3,2,6,5} \ , \
	  w_{1,2,3,8,5,4,3,6,5,4} \cr
	 &w_{1,2,3,8,5,4,3,6,5,8} \ , \
	  w_{1,2,4,5,6,8,5,4,3,6} \ , \
	  w_{1,2,4,5,6,8,5,4,6,5} \ , \
	  w_{1,2,4,5,6,8,5,4,6,7} \cr
	 &w_{1,2,4,5,8,5,4,3,2,1} \ , \
	  w_{1,2,4,5,8,5,4,3,2,6} \ , \
	  w_{1,2,4,5,8,5,4,3,6,5} \ , \
	  w_{1,2,4,5,8,5,4,3,6,7} \cr
	 &w_{1,2,4,5,8,5,4,6,5,7} \ , \
	  w_{1,2,4,5,8,5,4,6,5,8} \ , \
	  w_{1,2,4,7,8,5,4,3,2,1} \ , \
	  w_{1,2,4,7,8,5,4,3,2,6} \cr
	 &w_{1,2,4,7,8,5,4,3,6,7} \ , \
	  w_{1,2,4,8,5,4,3,2,1,6} \ , \
	  w_{1,2,4,8,5,4,3,2,6,5} \ , \
	  w_{1,2,4,8,5,4,3,2,6,7} \cr
	 &w_{1,2,4,8,5,4,3,6,5,4} \ , \
	  w_{1,2,4,8,5,4,3,6,5,7} \ , \
	  w_{1,2,4,8,5,4,3,6,5,8} \ , \
	  w_{1,2,4,8,5,4,6,5,7,8} \cr
	 &w_{1,2,8,5,4,3,2,1,6,5} \ , \
	  w_{1,2,8,5,4,3,2,6,5,4} \ , \
	  w_{1,2,8,5,4,3,2,6,5,8} \ , \
	  w_{1,2,8,5,4,3,6,5,4,8} \cr
	 &w_{1,4,3,8,5,4,3,2,1,6} \ , \
	  w_{1,4,3,8,5,4,3,2,6,5} \ , \
	  w_{1,4,3,8,5,4,3,2,6,7} \ , \
	  w_{1,4,3,8,5,4,3,6,5,4} \cr
	 &w_{1,4,3,8,5,4,3,6,5,7} \ , \
	  w_{1,4,3,8,5,4,6,5,7,6} \ , \
	  w_{1,4,5,6,8,5,4,6,5,7} \ , \
	  w_{1,4,5,6,8,5,4,6,5,8} \cr
	 &w_{1,4,5,8,5,4,3,2,1,6} \ , \
	  w_{1,4,5,8,5,4,3,2,6,5} \ , \
	  w_{1,4,5,8,5,4,3,2,6,7} \ , \
	  w_{1,4,5,8,5,4,3,6,5,7} \cr
	 &w_{1,4,5,8,5,4,3,6,5,8} \ , \
	  w_{1,4,5,8,5,4,6,5,7,6} \ , \
	  w_{1,4,5,8,5,4,6,5,7,8} \ , \
	  w_{1,4,7,8,5,4,3,2,1,6} \cr
	 &w_{1,4,7,8,5,4,3,2,6,7} \ , \
	  w_{1,4,8,5,4,3,2,1,6,5} \ , \
	  w_{1,4,8,5,4,3,2,1,6,7} \ , \
	  w_{1,4,8,5,4,3,2,6,5,4} \cr
	 &w_{1,4,8,5,4,3,2,6,5,7} \ , \
	  w_{1,4,8,5,4,3,2,6,5,8} \ , \
	  w_{1,4,8,5,4,3,6,5,4,7} \ , \
	  w_{1,4,8,5,4,3,6,5,4,8} \cr
	 &w_{1,4,8,5,4,3,6,5,7,8} \ , \
	  w_{1,8,5,4,3,2,1,6,5,4} \ , \
	  w_{1,8,5,4,3,2,1,6,5,8} \ , \
	  w_{1,8,5,4,3,2,6,5,4,3} \cr
	 &w_{1,8,5,4,3,2,6,5,4,8} \ , \
	  w_{1,8,5,4,3,6,5,4,8,5} \ , \
	  w_{4,3,2,8,5,4,3,2,1,6} \ , \
	  w_{4,3,2,8,5,4,3,2,6,5} \cr
	 &w_{4,3,2,8,5,4,3,2,6,7} \ , \
	  w_{4,3,2,8,5,4,3,6,5,4} \ , \
	  w_{4,3,2,8,5,4,3,6,5,7} \ , \
	  w_{4,3,2,8,5,4,6,5,7,6} \cr
	 &w_{4,3,8,5,4,3,2,1,6,5} \ , \
	  w_{4,3,8,5,4,3,2,1,6,7} \ , \
	  w_{4,3,8,5,4,3,2,6,5,4} \ , \
	  w_{4,3,8,5,4,3,2,6,5,7} \cr
	 &w_{4,3,8,5,4,3,6,5,4,7} \ , \
	  w_{4,3,8,5,4,3,6,5,7,6} \ , \
	  w_{4,5,6,8,5,4,6,5,7,6} \ , \
	  w_{4,5,6,8,5,4,6,5,7,8} \cr
	 &w_{4,5,8,5,4,3,2,1,6,5} \ , \
	  w_{4,5,8,5,4,3,2,1,6,7} \ , \
	  w_{4,5,8,5,4,3,2,6,5,7} \ , \
	  w_{4,5,8,5,4,3,2,6,5,8} \cr
	 &w_{4,5,8,5,4,3,6,5,7,6} \ , \
	  w_{4,5,8,5,4,3,6,5,7,8} \ , \
	  w_{4,5,8,5,4,6,5,7,6,8} \ , \
	  w_{4,7,8,5,4,3,2,1,6,7} \cr
	 &w_{4,8,5,4,3,2,1,6,5,4} \ , \
	  w_{4,8,5,4,3,2,1,6,5,7} \ , \
	  w_{4,8,5,4,3,2,1,6,5,8} \ , \
	  w_{4,8,5,4,3,2,6,5,4,3} \cr
	 &w_{4,8,5,4,3,2,6,5,4,7} \ , \
	  w_{4,8,5,4,3,2,6,5,4,8} \ , \
	  w_{4,8,5,4,3,2,6,5,7,8} \ , \
	  w_{4,8,5,4,3,6,5,4,7,8} \cr
	 &w_{8,5,4,3,2,1,6,5,4,3} \ , \
	  w_{8,5,4,3,2,1,6,5,4,8} \ , \
	  w_{8,5,4,3,2,6,5,4,3,8} \ , \
	  w_{8,5,4,3,2,6,5,4,8,5} \cr  &~}$ & \cr  \noalign{\hrule}
& & $\ell=11$ & &
$\eqalign{&~ \cr
	 &w_{1,2,3,8,5,4,3,2,1,6,5}  \ , \
	  w_{1,2,3,8,5,4,3,2,6,5,4}  \ , \
	  w_{1,2,3,8,5,4,3,2,6,5,8}  \ , \
	  w_{1,2,3,8,5,4,3,6,5,4,8}  \cr
	 &w_{1,2,4,5,6,8,5,4,3,6,5}  \ , \
	  w_{1,2,4,5,6,8,5,4,3,6,7}  \ , \
	  w_{1,2,4,5,6,8,5,4,6,5,7}  \ , \
	  w_{1,2,4,5,6,8,5,4,6,5,8}  \cr
	 &w_{1,2,4,5,8,5,4,3,2,1,6}  \ , \
	  w_{1,2,4,5,8,5,4,3,2,6,5}  \ , \
	  w_{1,2,4,5,8,5,4,3,2,6,7}  \ , \
	  w_{1,2,4,5,8,5,4,3,6,5,4}  \cr
	 &w_{1,2,4,5,8,5,4,3,6,5,7}  \ , \
	  w_{1,2,4,5,8,5,4,3,6,5,8}  \ , \
	  w_{1,2,4,5,8,5,4,6,5,7,6}  \ , \
	  w_{1,2,4,5,8,5,4,6,5,7,8}  \cr
	 &w_{1,2,4,7,8,5,4,3,2,1,6}  \ , \
	  w_{1,2,4,7,8,5,4,3,2,6,7}  \ , \
	  w_{1,2,4,8,5,4,3,2,1,6,5}  \ , \
	  w_{1,2,4,8,5,4,3,2,1,6,7}  \cr
	 &w_{1,2,4,8,5,4,3,2,6,5,4}  \ , \
	  w_{1,2,4,8,5,4,3,2,6,5,7}  \ , \
	  w_{1,2,4,8,5,4,3,2,6,5,8}  \ , \
	  w_{1,2,4,8,5,4,3,6,5,4,7}  \cr
	 &w_{1,2,4,8,5,4,3,6,5,4,8}  \ , \
	  w_{1,2,4,8,5,4,3,6,5,7,8}  \ , \
	  w_{1,2,8,5,4,3,2,1,6,5,4}  \ , \
	  w_{1,2,8,5,4,3,2,1,6,5,8}  \cr
	 &w_{1,2,8,5,4,3,2,6,5,4,3}  \ , \
	  w_{1,2,8,5,4,3,2,6,5,4,8}  \ , \
	  w_{1,2,8,5,4,3,6,5,4,8,5}  \ , \
	  w_{1,4,3,8,5,4,3,2,1,6,5}  \cr
	 &w_{1,4,3,8,5,4,3,2,1,6,7}  \ , \
	  w_{1,4,3,8,5,4,3,2,6,5,4}  \ , \
	  w_{1,4,3,8,5,4,3,2,6,5,7}  \ , \
	  w_{1,4,3,8,5,4,3,6,5,4,7}  \cr
	 &w_{1,4,3,8,5,4,3,6,5,7,6}  \ , \
	  w_{1,4,5,6,8,5,4,6,5,7,6}  \ , \
	  w_{1,4,5,6,8,5,4,6,5,7,8}  \ , \
	  w_{1,4,5,8,5,4,3,2,1,6,5}  \cr
	 &w_{1,4,5,8,5,4,3,2,1,6,7}  \ , \
	  w_{1,4,5,8,5,4,3,2,6,5,7}  \ , \
	  w_{1,4,5,8,5,4,3,2,6,5,8}  \ , \
	  w_{1,4,5,8,5,4,3,6,5,7,6}  \cr
	 &w_{1,4,5,8,5,4,3,6,5,7,8}  \ , \
	  w_{1,4,5,8,5,4,6,5,7,6,8}  \ , \
	  w_{1,4,7,8,5,4,3,2,1,6,7}  \ , \
	  w_{1,4,8,5,4,3,2,1,6,5,4}  \cr
	 &w_{1,4,8,5,4,3,2,1,6,5,7}  \ , \
	  w_{1,4,8,5,4,3,2,1,6,5,8}  \ , \
	  w_{1,4,8,5,4,3,2,6,5,4,3}  \ , \
	  w_{1,4,8,5,4,3,2,6,5,4,7}  \cr
	 &w_{1,4,8,5,4,3,2,6,5,4,8}  \ , \
	  w_{1,4,8,5,4,3,2,6,5,7,8}  \ , \
	  w_{1,4,8,5,4,3,6,5,4,7,8}  \ , \
	  w_{1,8,5,4,3,2,1,6,5,4,3}  \cr
	 &w_{1,8,5,4,3,2,1,6,5,4,8}  \ , \
	  w_{1,8,5,4,3,2,6,5,4,3,8}  \ , \
	  w_{1,8,5,4,3,2,6,5,4,8,5}  \ , \
	  w_{4,3,2,8,5,4,3,2,1,6,5}  \cr
	 &w_{4,3,2,8,5,4,3,2,1,6,7}  \ , \
	  w_{4,3,2,8,5,4,3,2,6,5,4}  \ , \
	  w_{4,3,2,8,5,4,3,2,6,5,7}  \ , \
	  w_{4,3,2,8,5,4,3,6,5,4,7}  \cr
	 &w_{4,3,2,8,5,4,3,6,5,7,6}  \ , \
	  w_{4,3,8,5,4,3,2,1,6,5,4}  \ , \
	  w_{4,3,8,5,4,3,2,1,6,5,7}  \ , \
	  w_{4,3,8,5,4,3,2,6,5,4,3}  \cr
	 &w_{4,3,8,5,4,3,2,6,5,4,7}  \ , \
	  w_{4,3,8,5,4,3,2,6,5,7,6}  \ , \
	  w_{4,3,8,5,4,3,6,5,4,7,6}  \ , \
	  w_{4,5,6,8,5,4,6,5,7,6,8}  \cr &~}$ & \cr  \noalign{\hrule}
}}$$

$$ \vbox{\tabskip=0pt \offinterlineskip
\halign to450pt{\strut#& \vrule#\tabskip=1em plus2em& \hfil#& \vrule#&
\hfil#\hfil& \vrule#\tabskip=0pt\cr \noalign{\hrule}
& &  & &
$\eqalign{&~ \cr
	 &w_{4,5,8,5,4,3,2,1,6,5,7}  \ , \
	  w_{4,5,8,5,4,3,2,1,6,5,8}  \ , \
	  w_{4,5,8,5,4,3,2,6,5,7,6}  \ , \
	  w_{4,5,8,5,4,3,2,6,5,7,8}  \cr
	 &w_{4,5,8,5,4,3,6,5,7,6,8}  \ , \
	  w_{4,8,5,4,3,2,1,6,5,4,3}  \ , \
	  w_{4,8,5,4,3,2,1,6,5,4,7}  \ , \
	  w_{4,8,5,4,3,2,1,6,5,4,8}  \cr
	 &w_{4,8,5,4,3,2,1,6,5,7,8}  \ , \
	  w_{4,8,5,4,3,2,6,5,4,3,7}  \ , \
	  w_{4,8,5,4,3,2,6,5,4,3,8}  \ , \
	  w_{4,8,5,4,3,2,6,5,4,7,8}  \cr
	 &w_{8,5,4,3,2,1,6,5,4,3,2}  \ , \
	  w_{8,5,4,3,2,1,6,5,4,3,8}  \ , \
	  w_{8,5,4,3,2,1,6,5,4,8,5}  \ , \
	  w_{8,5,4,3,2,6,5,4,3,8,5}  \cr  &~}$ & \cr  \noalign{\hrule}
& & $\ell=12$ & &
$\eqalign{&~ \cr
	 &w_{1,2,3,8,5,4,3,2,1,6,5,4} \ , \
	  w_{1,2,3,8,5,4,3,2,1,6,5,8} \ , \
	  w_{1,2,3,8,5,4,3,2,6,5,4,3} \cr
	 &w_{1,2,3,8,5,4,3,2,6,5,4,8} \ , \
	  w_{1,2,3,8,5,4,3,6,5,4,8,5} \ , \
	  w_{1,2,4,5,6,8,5,4,3,6,5,4} \cr
	 &w_{1,2,4,5,6,8,5,4,3,6,5,7} \ , \
	  w_{1,2,4,5,6,8,5,4,3,6,5,8} \ , \
	  w_{1,2,4,5,6,8,5,4,6,5,7,6} \cr
	 &w_{1,2,4,5,6,8,5,4,6,5,7,8} \ , \
	  w_{1,2,4,5,8,5,4,3,2,1,6,5} \ , \
	  w_{1,2,4,5,8,5,4,3,2,1,6,7} \cr
	 &w_{1,2,4,5,8,5,4,3,2,6,5,4} \ , \
	  w_{1,2,4,5,8,5,4,3,2,6,5,7} \ , \
	  w_{1,2,4,5,8,5,4,3,2,6,5,8} \cr
	 &w_{1,2,4,5,8,5,4,3,6,5,4,7} \ , \
	  w_{1,2,4,5,8,5,4,3,6,5,4,8} \ , \
	  w_{1,2,4,5,8,5,4,3,6,5,7,6} \cr
	 &w_{1,2,4,5,8,5,4,3,6,5,7,8} \ , \
	  w_{1,2,4,5,8,5,4,6,5,7,6,8} \ , \
	  w_{1,2,4,7,8,5,4,3,2,1,6,7} \cr
	 &w_{1,2,4,8,5,4,3,2,1,6,5,4} \ , \
	  w_{1,2,4,8,5,4,3,2,1,6,5,7} \ , \
	  w_{1,2,4,8,5,4,3,2,1,6,5,8} \cr
	 &w_{1,2,4,8,5,4,3,2,6,5,4,3} \ , \
	  w_{1,2,4,8,5,4,3,2,6,5,4,7} \ , \
	  w_{1,2,4,8,5,4,3,2,6,5,4,8} \cr
	 &w_{1,2,4,8,5,4,3,2,6,5,7,8} \ , \
	  w_{1,2,4,8,5,4,3,6,5,4,7,8} \ , \
	  w_{1,2,4,8,5,4,3,6,5,4,8,5} \cr
	 &w_{1,2,8,5,4,3,2,1,6,5,4,3} \ , \
	  w_{1,2,8,5,4,3,2,1,6,5,4,8} \ , \
	  w_{1,2,8,5,4,3,2,6,5,4,3,8} \cr
	 &w_{1,2,8,5,4,3,2,6,5,4,8,5} \ , \
	  w_{1,2,8,5,4,3,6,5,4,8,5,6} \ , \
	  w_{1,4,3,8,5,4,3,2,1,6,5,4} \cr
	 &w_{1,4,3,8,5,4,3,2,1,6,5,7} \ , \
	  w_{1,4,3,8,5,4,3,2,6,5,4,3} \ , \
	  w_{1,4,3,8,5,4,3,2,6,5,4,7} \cr
	 &w_{1,4,3,8,5,4,3,2,6,5,7,6} \ , \
	  w_{1,4,3,8,5,4,3,6,5,4,7,6} \ , \
	  w_{1,4,5,6,8,5,4,6,5,7,6,8} \cr
	 &w_{1,4,5,8,5,4,3,2,1,6,5,7} \ , \
	  w_{1,4,5,8,5,4,3,2,1,6,5,8} \ , \
	  w_{1,4,5,8,5,4,3,2,6,5,7,6} \cr
	 &w_{1,4,5,8,5,4,3,2,6,5,7,8} \ , \
	  w_{1,4,5,8,5,4,3,6,5,7,6,8} \ , \
	  w_{1,4,8,5,4,3,2,1,6,5,4,3} \cr
	 &w_{1,4,8,5,4,3,2,1,6,5,4,7} \ , \
	  w_{1,4,8,5,4,3,2,1,6,5,4,8} \ , \
	  w_{1,4,8,5,4,3,2,1,6,5,7,8} \cr
	 &w_{1,4,8,5,4,3,2,6,5,4,3,7} \ , \
	  w_{1,4,8,5,4,3,2,6,5,4,3,8} \ , \
	  w_{1,4,8,5,4,3,2,6,5,4,7,8} \cr
	 &w_{1,8,5,4,3,2,1,6,5,4,3,2} \ , \
	  w_{1,8,5,4,3,2,1,6,5,4,3,8} \ , \
	  w_{1,8,5,4,3,2,1,6,5,4,8,5} \cr
	 &w_{1,8,5,4,3,2,6,5,4,3,8,5} \ , \
	  w_{4,3,2,8,5,4,3,2,1,6,5,4} \ , \
	  w_{4,3,2,8,5,4,3,2,1,6,5,7} \cr
	 &w_{4,3,2,8,5,4,3,2,6,5,4,3} \ , \
	  w_{4,3,2,8,5,4,3,2,6,5,4,7} \ , \
	  w_{4,3,2,8,5,4,3,2,6,5,7,6} \cr
	 &w_{4,3,2,8,5,4,3,6,5,4,7,6} \ , \
	  w_{4,3,8,5,4,3,2,1,6,5,4,3} \ , \
	  w_{4,3,8,5,4,3,2,1,6,5,4,7} \cr
	 &w_{4,3,8,5,4,3,2,1,6,5,7,6} \ , \
	  w_{4,3,8,5,4,3,2,6,5,4,3,7} \ , \
	  w_{4,3,8,5,4,3,2,6,5,4,7,6} \cr
	 &w_{4,5,6,8,5,4,6,5,7,6,8,5} \ , \
	  w_{4,5,8,5,4,3,2,1,6,5,7,6} \ , \
	  w_{4,5,8,5,4,3,2,1,6,5,7,8} \cr
	 &w_{4,5,8,5,4,3,2,6,5,7,6,8} \ , \
	  w_{4,8,5,4,3,2,1,6,5,4,3,2} \ , \
	  w_{4,8,5,4,3,2,1,6,5,4,3,7} \cr
	 &w_{4,8,5,4,3,2,1,6,5,4,3,8} \ , \
	  w_{4,8,5,4,3,2,1,6,5,4,7,8} \ , \
	  w_{4,8,5,4,3,2,6,5,4,3,7,8} \cr
	 &w_{8,5,4,3,2,1,6,5,4,3,2,8} \ , \
	  w_{8,5,4,3,2,1,6,5,4,3,8,5} \ , \
	  w_{8,5,4,3,2,6,5,4,3,8,5,4} \cr     &~}$ & \cr  \noalign{\hrule}
& & $\ell=13$ & &
$\eqalign{&~ \cr
	 &w_{1,2,3,8,5,4,3,2,1,6,5,4,3}  \ , \
	  w_{1,2,3,8,5,4,3,2,1,6,5,4,8}  \ , \
	  w_{1,2,3,8,5,4,3,2,6,5,4,3,8}  \cr
	 &w_{1,2,3,8,5,4,3,2,6,5,4,8,5}  \ , \
	  w_{1,2,3,8,5,4,3,6,5,4,8,5,6}  \ , \
	  w_{1,2,4,5,6,8,5,4,3,6,5,4,7}  \cr
	 &w_{1,2,4,5,6,8,5,4,3,6,5,4,8}  \ , \
	  w_{1,2,4,5,6,8,5,4,3,6,5,7,6}  \ , \
	  w_{1,2,4,5,6,8,5,4,3,6,5,7,8}  \cr
	 &w_{1,2,4,5,6,8,5,4,6,5,7,6,8}  \ , \
	  w_{1,2,4,5,8,5,4,3,2,1,6,5,4}  \ , \
	  w_{1,2,4,5,8,5,4,3,2,1,6,5,7}  \cr
	 &w_{1,2,4,5,8,5,4,3,2,1,6,5,8}  \ , \
	  w_{1,2,4,5,8,5,4,3,2,6,5,4,7}  \ , \
	  w_{1,2,4,5,8,5,4,3,2,6,5,4,8}  \cr
	 &w_{1,2,4,5,8,5,4,3,2,6,5,7,6}  \ , \
	  w_{1,2,4,5,8,5,4,3,2,6,5,7,8}  \ , \
	  w_{1,2,4,5,8,5,4,3,6,5,4,7,6}  \cr
	 &w_{1,2,4,5,8,5,4,3,6,5,4,7,8}  \ , \
	  w_{1,2,4,5,8,5,4,3,6,5,4,8,5}  \ , \
	  w_{1,2,4,5,8,5,4,3,6,5,7,6,8}  \cr
	 &w_{1,2,4,8,5,4,3,2,1,6,5,4,3}  \ , \
	  w_{1,2,4,8,5,4,3,2,1,6,5,4,7}  \ , \
	  w_{1,2,4,8,5,4,3,2,1,6,5,4,8}  \cr
	 &w_{1,2,4,8,5,4,3,2,1,6,5,7,8}  \ , \
	  w_{1,2,4,8,5,4,3,2,6,5,4,3,7}  \ , \
	  w_{1,2,4,8,5,4,3,2,6,5,4,3,8}  \cr  &~}$ & \cr  \noalign{\hrule}
}}$$

$$ \vbox{\tabskip=0pt \offinterlineskip
\halign to450pt{\strut#& \vrule#\tabskip=1em plus2em& \hfil#& \vrule#&
\hfil#\hfil& \vrule#\tabskip=0pt\cr \noalign{\hrule}
& & & &
$\eqalign{&~ \cr
	 &w_{1,2,4,8,5,4,3,2,6,5,4,7,8}  \ , \
	  w_{1,2,4,8,5,4,3,2,6,5,4,8,5}  \ , \
	  w_{1,2,4,8,5,4,3,6,5,4,7,8,5}  \cr
	 &w_{1,2,4,8,5,4,3,6,5,4,8,5,6}  \ , \
	  w_{1,2,8,5,4,3,2,1,6,5,4,3,2}  \ , \
	  w_{1,2,8,5,4,3,2,1,6,5,4,3,8}  \cr
	 &w_{1,2,8,5,4,3,2,1,6,5,4,8,5}  \ , \
	  w_{1,2,8,5,4,3,2,6,5,4,3,8,5}  \ , \
	  w_{1,2,8,5,4,3,2,6,5,4,8,5,6}  \cr
	 &w_{1,2,8,5,4,3,6,5,4,8,5,6,7}  \ , \
	  w_{1,4,3,8,5,4,3,2,1,6,5,4,3}  \ , \
	  w_{1,4,3,8,5,4,3,2,1,6,5,4,7}  \cr
	 &w_{1,4,3,8,5,4,3,2,1,6,5,7,6}  \ , \
	  w_{1,4,3,8,5,4,3,2,6,5,4,3,7}  \ , \
	  w_{1,4,3,8,5,4,3,2,6,5,4,7,6}  \cr
	 &w_{1,4,5,6,8,5,4,6,5,7,6,8,5}  \ , \
	  w_{1,4,5,8,5,4,3,2,1,6,5,7,6}  \ , \
	  w_{1,4,5,8,5,4,3,2,1,6,5,7,8}  \cr
	 &w_{1,4,5,8,5,4,3,2,6,5,7,6,8}  \ , \
	  w_{1,4,8,5,4,3,2,1,6,5,4,3,2}  \ , \
	  w_{1,4,8,5,4,3,2,1,6,5,4,3,7}  \cr
	 &w_{1,4,8,5,4,3,2,1,6,5,4,3,8}  \ , \
	  w_{1,4,8,5,4,3,2,1,6,5,4,7,8}  \ , \
	  w_{1,4,8,5,4,3,2,6,5,4,3,7,8}  \cr
	 &w_{1,8,5,4,3,2,1,6,5,4,3,2,8}  \ , \
	  w_{1,8,5,4,3,2,1,6,5,4,3,8,5}  \ , \
	  w_{1,8,5,4,3,2,6,5,4,3,8,5,4}  \cr
	 &w_{4,3,2,8,5,4,3,2,1,6,5,4,3}  \ , \
	  w_{4,3,2,8,5,4,3,2,1,6,5,4,7}  \ , \
	  w_{4,3,2,8,5,4,3,2,1,6,5,7,6}  \cr
	 &w_{4,3,2,8,5,4,3,2,6,5,4,3,7}  \ , \
	  w_{4,3,2,8,5,4,3,2,6,5,4,7,6}  \ , \
	  w_{4,3,2,8,5,4,3,6,5,4,7,6,5}  \cr
	 &w_{4,3,8,5,4,3,2,1,6,5,4,3,2}  \ , \
	  w_{4,3,8,5,4,3,2,1,6,5,4,3,7}  \ , \
	  w_{4,3,8,5,4,3,2,1,6,5,4,7,6}  \cr
	 &w_{4,3,8,5,4,3,2,6,5,4,3,7,6}  \ , \
	  w_{4,5,8,5,4,3,2,1,6,5,7,6,8}  \ , \
	  w_{4,8,5,4,3,2,1,6,5,4,3,2,7}  \cr
	 &w_{4,8,5,4,3,2,1,6,5,4,3,2,8}  \ , \
	  w_{4,8,5,4,3,2,1,6,5,4,3,7,8}  \ , \
	  w_{8,5,4,3,2,1,6,5,4,3,2,8,5}  \cr
	 &w_{8,5,4,3,2,1,6,5,4,3,8,5,4}  \cr  &~}$ & \cr  \noalign{\hrule}
& & $\ell=14$ & &
$\eqalign{&~ \cr
	 &w_{1,2,3,8,5,4,3,2,1,6,5,4,3,2} \ , \
	  w_{1,2,3,8,5,4,3,2,1,6,5,4,3,8} \ , \
	  w_{1,2,3,8,5,4,3,2,1,6,5,4,8,5} \cr
	 &w_{1,2,3,8,5,4,3,2,6,5,4,3,8,5} \ , \
	  w_{1,2,3,8,5,4,3,2,6,5,4,8,5,6} \ , \
	  w_{1,2,3,8,5,4,3,6,5,4,8,5,6,7} \cr
	 &w_{1,2,4,5,6,8,5,4,3,6,5,4,7,6} \ , \
	  w_{1,2,4,5,6,8,5,4,3,6,5,4,7,8} \ , \
	  w_{1,2,4,5,6,8,5,4,3,6,5,4,8,5} \cr
	 &w_{1,2,4,5,6,8,5,4,3,6,5,7,6,8} \ , \
	  w_{1,2,4,5,6,8,5,4,6,5,7,6,8,5} \ , \
	  w_{1,2,4,5,8,5,4,3,2,1,6,5,4,7} \cr
	 &w_{1,2,4,5,8,5,4,3,2,1,6,5,4,8} \ , \
	  w_{1,2,4,5,8,5,4,3,2,1,6,5,7,6} \ , \
	  w_{1,2,4,5,8,5,4,3,2,1,6,5,7,8} \cr
	 &w_{1,2,4,5,8,5,4,3,2,6,5,4,7,6} \ , \
	  w_{1,2,4,5,8,5,4,3,2,6,5,4,7,8} \ , \
	  w_{1,2,4,5,8,5,4,3,2,6,5,4,8,5} \cr
	 &w_{1,2,4,5,8,5,4,3,2,6,5,7,6,8} \ , \
	  w_{1,2,4,5,8,5,4,3,6,5,4,7,6,5} \ , \
	  w_{1,2,4,5,8,5,4,3,6,5,4,7,6,8} \cr
	 &w_{1,2,4,5,8,5,4,3,6,5,4,7,8,5} \ , \
	  w_{1,2,4,5,8,5,4,3,6,5,4,8,5,6} \ , \
	  w_{1,2,4,8,5,4,3,2,1,6,5,4,3,2} \cr
	 &w_{1,2,4,8,5,4,3,2,1,6,5,4,3,7} \ , \
	  w_{1,2,4,8,5,4,3,2,1,6,5,4,3,8} \ , \
	  w_{1,2,4,8,5,4,3,2,1,6,5,4,7,8} \cr
	 &w_{1,2,4,8,5,4,3,2,1,6,5,4,8,5} \ , \
	  w_{1,2,4,8,5,4,3,2,6,5,4,3,7,8} \ , \
	  w_{1,2,4,8,5,4,3,2,6,5,4,3,8,5} \cr
	 &w_{1,2,4,8,5,4,3,2,6,5,4,7,8,5} \ , \
	  w_{1,2,4,8,5,4,3,2,6,5,4,8,5,6} \ , \
	  w_{1,2,4,8,5,4,3,6,5,4,7,8,5,6} \cr
	 &w_{1,2,4,8,5,4,3,6,5,4,8,5,6,7} \ , \
	  w_{1,2,8,5,4,3,2,1,6,5,4,3,2,8} \ , \
	  w_{1,2,8,5,4,3,2,1,6,5,4,3,8,5} \cr
	 &w_{1,2,8,5,4,3,2,1,6,5,4,8,5,6} \ , \
	  w_{1,2,8,5,4,3,2,6,5,4,3,8,5,4} \ , \
	  w_{1,2,8,5,4,3,2,6,5,4,3,8,5,6} \cr
	 &w_{1,2,8,5,4,3,2,6,5,4,8,5,6,7} \ , \
	  w_{1,4,3,8,5,4,3,2,1,6,5,4,3,2} \ , \
	  w_{1,4,3,8,5,4,3,2,1,6,5,4,3,7} \cr
	 &w_{1,4,3,8,5,4,3,2,1,6,5,4,7,6} \ , \
	  w_{1,4,3,8,5,4,3,2,6,5,4,3,7,6} \ , \
	  w_{1,4,5,8,5,4,3,2,1,6,5,7,6,8} \cr
	 &w_{1,4,8,5,4,3,2,1,6,5,4,3,2,7} \ , \
	  w_{1,4,8,5,4,3,2,1,6,5,4,3,2,8} \ , \
	  w_{1,4,8,5,4,3,2,1,6,5,4,3,7,8} \cr
	 &w_{1,8,5,4,3,2,1,6,5,4,3,2,8,5} \ , \
	  w_{1,8,5,4,3,2,1,6,5,4,3,8,5,4} \ , \
	  w_{4,3,2,8,5,4,3,2,1,6,5,4,3,2} \cr
	 &w_{4,3,2,8,5,4,3,2,1,6,5,4,3,7} \ , \
	  w_{4,3,2,8,5,4,3,2,1,6,5,4,7,6} \ , \
	  w_{4,3,2,8,5,4,3,2,6,5,4,3,7,6} \cr
	 &w_{4,3,2,8,5,4,3,2,6,5,4,7,6,5} \ , \
	  w_{4,3,2,8,5,4,3,6,5,4,7,6,5,8} \ , \
	  w_{4,3,8,5,4,3,2,1,6,5,4,3,2,7} \cr
	 &w_{4,3,8,5,4,3,2,1,6,5,4,3,7,6} \ , \
	  w_{4,8,5,4,3,2,1,6,5,4,3,2,7,8} \ , \
	  w_{8,5,4,3,2,1,6,5,4,3,2,8,5,4} \cr &~}$ & \cr  \noalign{\hrule}
& & $\ell=15$ & &
$\eqalign{&~ \cr
	 &w_{1,2,3,8,5,4,3,2,1,6,5,4,3,2,8}  \ , \
	  w_{1,2,3,8,5,4,3,2,1,6,5,4,3,8,5}  \ , \
	  w_{1,2,3,8,5,4,3,2,1,6,5,4,8,5,6}  \cr
	 &w_{1,2,3,8,5,4,3,2,6,5,4,3,8,5,4}  \ , \
	  w_{1,2,3,8,5,4,3,2,6,5,4,3,8,5,6}  \ , \
	  w_{1,2,3,8,5,4,3,2,6,5,4,8,5,6,7}  \cr
	 &w_{1,2,4,5,6,8,5,4,3,6,5,4,7,6,5}  \ , \
	  w_{1,2,4,5,6,8,5,4,3,6,5,4,7,6,8}  \ , \
	  w_{1,2,4,5,6,8,5,4,3,6,5,4,7,8,5}  \cr
	 &w_{1,2,4,5,6,8,5,4,3,6,5,4,8,5,6}  \ , \
	  w_{1,2,4,5,6,8,5,4,3,6,5,7,6,8,5}  \ , \
	  w_{1,2,4,5,8,5,4,3,2,1,6,5,4,7,6}  \cr
	 &w_{1,2,4,5,8,5,4,3,2,1,6,5,4,7,8}  \ , \
	  w_{1,2,4,5,8,5,4,3,2,1,6,5,4,8,5}  \ , \
	  w_{1,2,4,5,8,5,4,3,2,1,6,5,7,6,8}  \cr  &~}$ & \cr  \noalign{\hrule}
}}$$

$$ \vbox{\tabskip=0pt \offinterlineskip
\halign to450pt{\strut#& \vrule#\tabskip=1em plus2em& \hfil#& \vrule#&
\hfil#\hfil& \vrule#\tabskip=0pt\cr \noalign{\hrule}
& & & &
$\eqalign{&~ \cr
	 &w_{1,2,4,5,8,5,4,3,2,6,5,4,7,6,5}  \ , \
	  w_{1,2,4,5,8,5,4,3,2,6,5,4,7,6,8}  \ , \
	  w_{1,2,4,5,8,5,4,3,2,6,5,4,7,8,5}  \cr
	 &w_{1,2,4,5,8,5,4,3,2,6,5,4,8,5,6}  \ , \
	  w_{1,2,4,5,8,5,4,3,6,5,4,7,6,5,8}  \ , \
	  w_{1,2,4,5,8,5,4,3,6,5,4,7,6,8,5}  \cr
	 &w_{1,2,4,5,8,5,4,3,6,5,4,7,8,5,6}  \ , \
	  w_{1,2,4,5,8,5,4,3,6,5,4,8,5,6,7}  \ , \
	  w_{1,2,4,8,5,4,3,2,1,6,5,4,3,2,7}  \cr
	 &w_{1,2,4,8,5,4,3,2,1,6,5,4,3,2,8}  \ , \
	  w_{1,2,4,8,5,4,3,2,1,6,5,4,3,7,8}  \ , \
	  w_{1,2,4,8,5,4,3,2,1,6,5,4,3,8,5}  \cr
	 &w_{1,2,4,8,5,4,3,2,1,6,5,4,7,8,5}  \ , \
	  w_{1,2,4,8,5,4,3,2,1,6,5,4,8,5,6}  \ , \
	  w_{1,2,4,8,5,4,3,2,6,5,4,3,7,8,5}  \cr
	 &w_{1,2,4,8,5,4,3,2,6,5,4,3,8,5,6}  \ , \
	  w_{1,2,4,8,5,4,3,2,6,5,4,7,8,5,6}  \ , \
	  w_{1,2,4,8,5,4,3,2,6,5,4,8,5,6,7}  \cr
	 &w_{1,2,4,8,5,4,3,6,5,4,7,8,5,6,7}  \ , \
	  w_{1,2,8,5,4,3,2,1,6,5,4,3,2,8,5}  \ , \
	  w_{1,2,8,5,4,3,2,1,6,5,4,3,8,5,4}  \cr
	 &w_{1,2,8,5,4,3,2,1,6,5,4,3,8,5,6}  \ , \
	  w_{1,2,8,5,4,3,2,1,6,5,4,8,5,6,7}  \ , \
	  w_{1,2,8,5,4,3,2,6,5,4,3,8,5,4,6}  \cr
	 &w_{1,2,8,5,4,3,2,6,5,4,3,8,5,6,7}  \ , \
	  w_{1,4,3,8,5,4,3,2,1,6,5,4,3,2,7}  \ , \
	  w_{1,4,3,8,5,4,3,2,1,6,5,4,3,7,6}  \cr
	 &w_{1,4,8,5,4,3,2,1,6,5,4,3,2,7,8}  \ , \
	  w_{1,8,5,4,3,2,1,6,5,4,3,2,8,5,4}  \ , \
	  w_{4,3,2,8,5,4,3,2,1,6,5,4,3,2,7}  \cr
	 &w_{4,3,2,8,5,4,3,2,1,6,5,4,3,7,6}  \ , \
	  w_{4,3,2,8,5,4,3,2,1,6,5,4,7,6,5}  \ , \
	  w_{4,3,2,8,5,4,3,2,6,5,4,3,7,6,5}  \cr
	 &w_{4,3,2,8,5,4,3,2,6,5,4,7,6,5,8}  \ , \
	  w_{4,3,8,5,4,3,2,1,6,5,4,3,2,7,6}  \ , \
	  w_{8,5,4,3,2,1,6,5,4,3,2,8,5,4,3}  \cr &~}$ & \cr  \noalign{\hrule}
& & $\ell=16$ & &
$\eqalign{&~ \cr
	 &w_{1,2,3,8,5,4,3,2,1,6,5,4,3,2,8,5} \ , \
	  w_{1,2,3,8,5,4,3,2,1,6,5,4,3,8,5,4} \ , \
	  w_{1,2,3,8,5,4,3,2,1,6,5,4,3,8,5,6} \cr
	 &w_{1,2,3,8,5,4,3,2,1,6,5,4,8,5,6,7} \ , \
	  w_{1,2,3,8,5,4,3,2,6,5,4,3,8,5,4,6} \ , \
	  w_{1,2,3,8,5,4,3,2,6,5,4,3,8,5,6,7} \cr
	 &w_{1,2,4,5,6,8,5,4,3,6,5,4,7,6,5,8} \ , \
	  w_{1,2,4,5,6,8,5,4,3,6,5,4,7,6,8,5} \ , \
	  w_{1,2,4,5,6,8,5,4,3,6,5,4,7,8,5,6} \cr
	 &w_{1,2,4,5,6,8,5,4,3,6,5,4,8,5,6,7} \ , \
	  w_{1,2,4,5,6,8,5,4,3,6,5,7,6,8,5,4} \ , \
	  w_{1,2,4,5,8,5,4,3,2,1,6,5,4,7,6,5} \cr
	 &w_{1,2,4,5,8,5,4,3,2,1,6,5,4,7,6,8} \ , \
	  w_{1,2,4,5,8,5,4,3,2,1,6,5,4,7,8,5} \ , \
	  w_{1,2,4,5,8,5,4,3,2,1,6,5,4,8,5,6} \cr
	 &w_{1,2,4,5,8,5,4,3,2,6,5,4,7,6,5,8} \ , \
	  w_{1,2,4,5,8,5,4,3,2,6,5,4,7,6,8,5} \ , \
	  w_{1,2,4,5,8,5,4,3,2,6,5,4,7,8,5,6} \cr
	 &w_{1,2,4,5,8,5,4,3,2,6,5,4,8,5,6,7} \ , \
	  w_{1,2,4,5,8,5,4,3,6,5,4,7,6,5,8,5} \ , \
	  w_{1,2,4,5,8,5,4,3,6,5,4,7,6,8,5,6} \cr
	 &w_{1,2,4,5,8,5,4,3,6,5,4,7,8,5,6,7} \ , \
	  w_{1,2,4,8,5,4,3,2,1,6,5,4,3,2,7,8} \ , \
	  w_{1,2,4,8,5,4,3,2,1,6,5,4,3,2,8,5} \cr
	 &w_{1,2,4,8,5,4,3,2,1,6,5,4,3,7,8,5} \ , \
	  w_{1,2,4,8,5,4,3,2,1,6,5,4,3,8,5,6} \ , \
	  w_{1,2,4,8,5,4,3,2,1,6,5,4,7,8,5,6} \cr
	 &w_{1,2,4,8,5,4,3,2,1,6,5,4,8,5,6,7} \ , \
	  w_{1,2,4,8,5,4,3,2,6,5,4,3,7,8,5,6} \ , \
	  w_{1,2,4,8,5,4,3,2,6,5,4,3,8,5,6,7} \cr
	 &w_{1,2,4,8,5,4,3,2,6,5,4,7,8,5,6,7} \ , \
	  w_{1,2,8,5,4,3,2,1,6,5,4,3,2,8,5,4} \ , \
	  w_{1,2,8,5,4,3,2,1,6,5,4,3,2,8,5,6} \cr
	 &w_{1,2,8,5,4,3,2,1,6,5,4,3,8,5,4,6} \ , \
	  w_{1,2,8,5,4,3,2,1,6,5,4,3,8,5,6,7} \ , \
	  w_{1,2,8,5,4,3,2,6,5,4,3,8,5,4,6,7} \cr
	 &w_{1,4,3,8,5,4,3,2,1,6,5,4,3,2,7,6} \ , \
	  w_{1,8,5,4,3,2,1,6,5,4,3,2,8,5,4,3} \ , \
	  w_{4,3,2,8,5,4,3,2,1,6,5,4,3,2,7,6} \cr
	 &w_{4,3,2,8,5,4,3,2,1,6,5,4,3,7,6,5} \ , \
	  w_{4,3,2,8,5,4,3,2,1,6,5,4,7,6,5,8} \ , \
	  w_{4,3,2,8,5,4,3,2,6,5,4,3,7,6,5,8} \cr &~}$ & \cr  \noalign{\hrule}
& & $\ell=17$ & &
$\eqalign{&~ \cr
	 &w_{1,2,3,8,5,4,3,2,1,6,5,4,3,2,8,5,4}  \ , \
	  w_{1,2,3,8,5,4,3,2,1,6,5,4,3,2,8,5,6}  \ , \
	  w_{1,2,3,8,5,4,3,2,1,6,5,4,3,8,5,4,6}  \cr
	 &w_{1,2,3,8,5,4,3,2,1,6,5,4,3,8,5,6,7}  \ , \
	  w_{1,2,3,8,5,4,3,2,6,5,4,3,8,5,4,6,5}  \ , \
	  w_{1,2,3,8,5,4,3,2,6,5,4,3,8,5,4,6,7}  \cr
	 &w_{1,2,4,5,6,8,5,4,3,6,5,4,7,6,5,8,5}  \ , \
	  w_{1,2,4,5,6,8,5,4,3,6,5,4,7,6,8,5,4}  \ , \
	  w_{1,2,4,5,6,8,5,4,3,6,5,4,7,6,8,5,6}  \cr
	 &w_{1,2,4,5,6,8,5,4,3,6,5,4,7,8,5,6,7}  \ , \
	  w_{1,2,4,5,8,5,4,3,2,1,6,5,4,7,6,5,8}  \ , \
	  w_{1,2,4,5,8,5,4,3,2,1,6,5,4,7,6,8,5}  \cr
	 &w_{1,2,4,5,8,5,4,3,2,1,6,5,4,7,8,5,6}  \ , \
	  w_{1,2,4,5,8,5,4,3,2,1,6,5,4,8,5,6,7}  \ , \
	  w_{1,2,4,5,8,5,4,3,2,6,5,4,7,6,5,8,5}  \cr
	 &w_{1,2,4,5,8,5,4,3,2,6,5,4,7,6,8,5,6}  \ , \
	  w_{1,2,4,5,8,5,4,3,2,6,5,4,7,8,5,6,7}  \ , \
	  w_{1,2,4,5,8,5,4,3,6,5,4,7,6,5,8,5,6}  \cr
	 &w_{1,2,4,5,8,5,4,3,6,5,4,7,6,8,5,6,7}  \ , \
	  w_{1,2,4,8,5,4,3,2,1,6,5,4,3,2,7,8,5}  \ , \
	  w_{1,2,4,8,5,4,3,2,1,6,5,4,3,2,8,5,6}  \cr
	 &w_{1,2,4,8,5,4,3,2,1,6,5,4,3,7,8,5,6}  \ , \
	  w_{1,2,4,8,5,4,3,2,1,6,5,4,3,8,5,6,7}  \ , \
	  w_{1,2,4,8,5,4,3,2,1,6,5,4,7,8,5,6,7}  \cr
	 &w_{1,2,4,8,5,4,3,2,6,5,4,3,7,8,5,6,7}  \ , \
	  w_{1,2,8,5,4,3,2,1,6,5,4,3,2,8,5,4,3}  \ , \
	  w_{1,2,8,5,4,3,2,1,6,5,4,3,2,8,5,4,6}  \cr
	 &w_{1,2,8,5,4,3,2,1,6,5,4,3,2,8,5,6,7}  \ , \
	  w_{1,2,8,5,4,3,2,1,6,5,4,3,8,5,4,6,7}  \ , \
	  w_{4,3,2,8,5,4,3,2,1,6,5,4,3,2,7,6,5}  \cr
	 &w_{4,3,2,8,5,4,3,2,1,6,5,4,3,7,6,5,8}  \cr  &~}$ & \cr  \noalign{\hrule}
}}$$

$$ \vbox{\tabskip=0pt \offinterlineskip
\halign to465pt{\strut#& \vrule#\tabskip=1em plus2em& \hfil#& \vrule#&
\hfil#\hfil& \vrule#\tabskip=0pt\cr \noalign{\hrule}
& &$\ell=18$  & &
$\eqalign{&~ \cr
	 &w_{1,2,3,8,5,4,3,2,1,6,5,4,3,2,8,5,4,3}  \ , \
	  w_{1,2,3,8,5,4,3,2,1,6,5,4,3,2,8,5,4,6}  \ , \
	  w_{1,2,3,8,5,4,3,2,1,6,5,4,3,2,8,5,6,7}  \cr
	 &w_{1,2,3,8,5,4,3,2,1,6,5,4,3,8,5,4,6,5}  \ , \
	  w_{1,2,3,8,5,4,3,2,1,6,5,4,3,8,5,4,6,7}  \ , \
	  w_{1,2,3,8,5,4,3,2,6,5,4,3,8,5,4,6,5,7}  \cr
	 &w_{1,2,4,5,6,8,5,4,3,6,5,4,7,6,5,8,5,4}  \ , \
	  w_{1,2,4,5,6,8,5,4,3,6,5,4,7,6,5,8,5,6}  \ , \
	  w_{1,2,4,5,6,8,5,4,3,6,5,4,7,6,8,5,4,6}  \cr
	 &w_{1,2,4,5,6,8,5,4,3,6,5,4,7,6,8,5,6,7}  \ , \
	  w_{1,2,4,5,8,5,4,3,2,1,6,5,4,7,6,5,8,5}  \ , \
	  w_{1,2,4,5,8,5,4,3,2,1,6,5,4,7,6,8,5,6}  \cr
	 &w_{1,2,4,5,8,5,4,3,2,1,6,5,4,7,8,5,6,7}  \ , \
	  w_{1,2,4,5,8,5,4,3,2,6,5,4,8,5,6,7,6,5}  \ , \
	  w_{1,2,4,5,8,5,4,3,2,6,5,4,8,5,7,6,5,8}  \cr
	 &w_{1,2,4,5,8,5,4,3,6,5,4,7,6,5,8,5,6,7}  \ , \
	  w_{1,2,4,8,5,4,3,2,1,6,5,4,3,2,7,8,5,6}  \ , \
	  w_{1,2,4,8,5,4,3,2,1,6,5,4,3,2,8,5,6,7}  \cr
	 &w_{1,2,4,8,5,4,3,2,1,6,5,4,3,7,8,5,6,7}  \ , \
	  w_{1,2,8,5,4,3,2,1,6,5,4,3,2,8,5,4,3,6}  \ , \
	  w_{1,2,8,5,4,3,2,1,6,5,4,3,2,8,5,4,6,7}  \cr
	 &w_{4,3,2,8,5,4,3,2,1,6,5,4,3,2,7,6,5,8}  \cr  &~}$ & \cr  \noalign{\hrule}
& &$\ell=19$  & &
$\eqalign{&~ \cr
	 &w_{1,2,3,8,5,4,3,2,1,6,5,4,3,2,8,5,4,3,6} \ , \
	  w_{1,2,3,8,5,4,3,2,1,6,5,4,3,2,8,5,4,6,5} \cr
	 &w_{1,2,3,8,5,4,3,2,1,6,5,4,3,2,8,5,4,6,7} \ , \
	  w_{1,2,3,8,5,4,3,2,1,6,5,4,3,8,5,4,6,5,7} \cr
	 &w_{1,2,3,8,5,4,3,2,6,5,4,3,8,5,4,6,5,7,6} \ , \
	  w_{1,2,4,5,6,8,5,4,3,6,5,4,7,6,5,8,5,4,6} \cr
	 &w_{1,2,4,5,6,8,5,4,3,6,5,4,7,6,5,8,5,6,7} \ , \
	  w_{1,2,4,5,6,8,5,4,3,6,5,4,7,6,8,5,4,6,7} \cr
	 &w_{1,2,4,5,8,5,4,3,2,1,6,5,4,7,6,5,8,5,6} \ , \
	  w_{1,2,4,5,8,5,4,3,2,1,6,5,4,7,6,8,5,6,7} \cr
	 &w_{1,2,4,5,8,5,4,3,2,6,5,4,8,5,6,7,6,5,8} \ , \
	  w_{1,2,4,8,5,4,3,2,1,6,5,4,3,2,7,8,5,6,7} \cr
	 &w_{1,2,8,5,4,3,2,1,6,5,4,3,2,8,5,4,3,6,7}  \cr  &~}$ & \cr  \noalign{\hrule}
& &$\ell=20$  & &
$\eqalign{&~ \cr
	 &w_{1,2,3,8,5,4,3,2,1,6,5,4,3,2,8,5,4,3,6,5} \ , \
	  w_{1,2,3,8,5,4,3,2,1,6,5,4,3,2,8,5,4,3,6,7} \cr
	 &w_{1,2,3,8,5,4,3,2,1,6,5,4,3,2,8,5,4,6,5,7} \ , \
	  w_{1,2,3,8,5,4,3,2,1,6,5,4,3,8,5,4,6,5,7,6} \cr
	 &w_{1,2,4,5,6,8,5,4,3,6,5,4,7,6,5,8,5,4,6,5} \ , \
	  w_{1,2,4,5,6,8,5,4,3,6,5,4,7,6,5,8,5,4,6,7} \cr
	 &w_{1,2,4,5,8,5,4,3,2,1,6,5,4,7,6,5,8,5,6,7}   \cr  &~}$ & \cr  \noalign{\hrule}
& &$\ell=21$  & &
$\eqalign{&~ \cr
	 &w_{1,2,3,8,5,4,3,2,1,6,5,4,3,2,8,5,4,3,6,5,4} \ , \
	  w_{1,2,3,8,5,4,3,2,1,6,5,4,3,2,8,5,4,3,6,5,7} \cr
	 &w_{1,2,3,8,5,4,3,2,1,6,5,4,3,2,8,5,4,6,5,7,6} \ , \
	  w_{1,2,4,5,6,8,5,4,3,6,5,4,7,6,5,8,5,4,6,5,7} \cr  &~}$ & \cr  \noalign{\hrule}
& &$\ell=22$  & &
$\eqalign{&~ \cr
	 &w_{1,2,3,8,5,4,3,2,1,6,5,4,3,2,8,5,4,3,6,5,4,7} \ , \
	  w_{1,2,3,8,5,4,3,2,1,6,5,4,3,2,8,5,4,3,6,5,7,6} \cr
	 &w_{1,2,4,5,6,8,5,4,3,6,5,4,7,6,5,8,5,4,6,5,7,6} \cr &~}$ & \cr  \noalign{\hrule}
& &$\ell=23$  & &
$\eqalign{&~ \cr
	 &w_{1,2,3,8,5,4,3,2,1,6,5,4,3,2,8,5,4,3,6,5,4,7,6} \cr &~}$ & \cr  \noalign{\hrule}
& &$\ell=24$  & &
$\eqalign{&~ \cr
	&w_{1,2,3,8,5,4,3,2,1,6,5,4,3,2,8,5,4,3,6,5,4,7,6,5} \cr &~}$ & \cr  \noalign{\hrule}
& &$\ell=25$  & &
$\eqalign{&~ \cr
	&w_{1,2,3,8,5,4,3,2,1,6,5,4,3,2,8,5,4,3,6,5,4,7,6,5,8} \cr &~}$ & \cr  \noalign{\hrule}
}}$$

\end